\documentclass{aastex61}
\usepackage[utf8]{inputenc}
\usepackage{nameref}
\usepackage{wasysym}
\usepackage{makecell}
\usepackage[caption=false]{subfig}
\usepackage{hyperref}
\usepackage[capitalise]{cleveref}

\graphicspath{{img/}}

\received{December 20, 2017}
\revised{January 8, 2018}
\submitjournal{ApJ}

\begin{document}
	
\title{On the diversity in mass and orbital radius \\ of giant planets formed via disk instability}
\author{Simon Müller}
\affiliation{Center for Theoretical Astrophysics and Cosmology \\
			Institute for Computational Science, University of Zürich \\
			Winterthurerstrasse 190, 8057 Zürich, Switzerland}
\email{muesim@physik.uzh.ch}

\author{Ravit Helled}
\affiliation{Center for Theoretical Astrophysics and Cosmology \\
			Institute for Computational Science, University of Zürich \\
			Winterthurerstrasse 190, 8057 Zürich, Switzerland}

\author{Lucio Mayer}
\affiliation{Center for Theoretical Astrophysics and Cosmology \\
			Institute for Computational Science, University of Zürich \\
			Winterthurerstrasse 190, 8057 Zürich, Switzerland}

\correspondingauthor{Simon Müller}

\begin{abstract}
We present a semi-analytical population synthesis model of protoplanetary clumps formed by disk instability at radial distances of 80 - 120 AU. Various clump density profiles, initial mass functions, protoplanetary disk models, stellar masses, and gap opening criteria are considered. When we use more realistic gap opening criteria, we find that gaps open only rarely, which strongly affects clump survival rates and their physical properties (mass, radius and radial distance). The inferred surviving population is then shifted towards less massive clumps at smaller radial distances. We also find that populations of surviving clumps are very sensitive to the model assumptions and used parameters. Depending on the chosen parameters, the protoplanets occupy a mass range between 0.01 and 16 $M_{\text{J}}$ and may either orbit close to the central star or as far out as 75 AU, with a sweet spot at 10-30 AU for the massive ones. However, in all the cases we consider we find that massive giant planets at very large radial distances are rare in qualitative agreement with current direct imaging surveys. We conclude that caution should be taken in deriving population synthesis models as well as when comparing the models' results with observations.

\end{abstract}


\section{Introduction} \label{sec:introduction}
The number of detected exoplanets and their characterization is proceeding at a quick pace - by early 2018,
thousands of confirmed exoplanets have been detected. This spectacular advancement allows a 
direct comparison of observations with the predictions of population synthesis models.
Although the statistical knowledge of the exoplanet population is confined by the currently accessible 
parameter space of detection techniques, this distribution is diverse and different to what one might 
naively expect from our Solar System. The variety exhibited by these objects provides a proving ground for the 
proposed theoretical frameworks of planet formation. 

There are two competing theories for giant planet formation: core accretion (CA) \citep{Mizuno1980,Pollack1996}, which is the standard one, and disk instability (DI) \citep{Cameron1978,Boss1997}. 
In the first model the formation of a giant planet begins with assembling a heavy-element core, which at some threshold mass leads to runaway gas accretion. Several population synthesis models for CA have been developed and refined over the years (see e.g. \citet{Ida2004,Ida2008,Mordasini2009,Mordasini2012,Ida2013,Benz2014}). 
These models provide a rather consistent agreement between theory and observations, and although the models can improve further, it is fair to say that they  have reached maturity. 
\par

While the CA model is consistent with many observations and can explain the formation of various planetary types in one theoretical framework, it also has weaknesses such as type 1 migration, formation of giant planets around metal-poor stars, the formation of very massive giant planets very fast and at large radial distances, etc (see \citet{Helled2013} for a review). These weaknesses have led to the revival of the DI  model where giant planets form as a result of fragmentation of a massive protoplanetary disk around a young star. This formation scenario might not be the standard one for giant planet formation, but it can operate in several conditions where CA fails. 
\par

Key advantages of the DI model are discussed in \citet{Helled2013}. Depending on the combination of the dominant processes, disk fragmentation can lead to various different outcomes. Gas giants can form with and without cores 
and be either metal-rich or metal-poor, and their mass can vary from the brown-dwarf regime to that of intermediate-mass planets. The expected frequent tidal disruption of clumps could explain the outburst activity in young protoplanetary disks. In addition, in the DI model, planet formation can occur in the early stages of disk evolution and at large radii. Additionally, it may also occur  in low metal environments, which is consistent with the planetary system HR 8799 \citep{Marois2008}. Clearly, this model also has disadvantages. First, it is unclear whether clumps survive to become gravitationally bound protoplanets; both rapid inward migration and tidal disruption work against their survival. Second, the formation of terrestrial planets is not easily explained, and third, this model cannot naturally predict the correlation of giant planet occurrence and stellar metallicity. Nevertheless, the possibility of forming giant planets by disk instability is intriguing and more research in this direction is desirable, in particular to establish a more coherent picture of the predicted population of planets in this formation scenario. 
 \par 

Gravitational instabilities develop when protoplanetary disks become unstable due to their own self-gravity and develop spiral density waves that lead to fragmentation. 
The mechanism relies on the disk being massive and cold enough, where the condition for disk instability can be given by the Toomre parameter \citep{Toomre1964}

\begin{equation}
Q = \frac{c_{\text{s}} \kappa}{\pi \text{G} \Sigma} < 1
\label{eq:ToomreQ}
\end{equation}

where $c_{\text{s}}$, $\kappa$ and $\Sigma$ are the local sound speed, epicyclic frequency and disk surface density. This expression captures the essential physics, namely that gas pressure and orbital motion tend to counteract the destabilization of the self-gravity of the disk. In the case of protoplanetary disks, self-gravitating clumps can form in spiral arms if $Q_{\text{min}} < 1.4$ \citep{Durisen2006} provided that cooling is efficient. 

In terms of population synthesis models, less work has been dedicated to the DI scenario and the outcome of this formation model is less certain although work in that direction has begun (e.g., \citet{Forgan2013,Forgan2015,Rice2015,Nayakshin2015b,Nayakshin2015c,Nayakshin2016,Forgan2017}. See section \ref{sec:comparison} for details). At the moment, in contrast to the CA population synthesis models, there are more inconsistencies in population synthesis models in the DI scenario, and the physical processes that dominate the final population are still being debated.
In this paper we explore the sensitivity of the inferred population in the DI scenario for different model assumptions using a simple population synthesis model. This model  builds on a 1D model to simulate the pre-collapse evolution of clumps embedded in a protoplanetary disk. A relatively simple model allows us to explore a large parameter space of variables involved, in particular, to evolve clumps using different assumptions and investigate the physical properties of surviving protoplanets. 

\section{Methods} \label{sec:methods}
The DI model makes distinct predictions for the very initial structure and formation epoch of giant planets. 
The most direct observational constraint would be to measure \textsl{when} and \textsl{where} giant planets form. 
Giants planets form rapidly (a few dynamical timescales) within disks that develop spiral arms and fragment. 
Therefore, the planet formation epoch should occur early in the disk's lifetime (a few $10^5$ years) and at large radii (where rapid cooling is expected). For a solar mass star, the ideal location for clump formation is at $\sim 10^{2}$ AU \citep{Clarke2009,Matzner2005,Rafikov2009}. While the initial mass of the protoplanets formed via disk instability is not well-constrained, it is estimated to be in the range of $1 - 10 M_{\text{J}}$ \citep{Forgan2011a}. 
\par 

To study the populations of protoplanetary clumps we build on the 1D semi-analytical model of \citet{Galvagni2014} which are based on fully 3D collapse simulations of clumps including varied thermodynamics. 
The clumps are assumed to have already formed by disk instability and are then evolved
inside the protoplanetary disk. For simplicity, we do not consider dynamical interactions between the clumps which can lead to ejection \citep{Terquem2002}. Instead, we allow for a wide range of initial conditions and include the following evolutionary processes: migration, mass accretion, tidal disruption, disk viscosity, and gap opening. 
The clump is then followed until it reaches second core collapse, i.e., when dissociation of molecular hydrogen initiates. At this point the clump shrinks and becomes denser by an order of magnitude \citep{Masunaga1998} and is protected from tidal disruptions and is likely to survive. We stress that \textit{our model does not make predictions of the actual population of planets}, but rather we form protoplanets that would still evolve further.

\subsection{Disk models} \label{sec:disk_models}
\begin{figure}[h]
	\centering
	\includegraphics[scale=0.32]{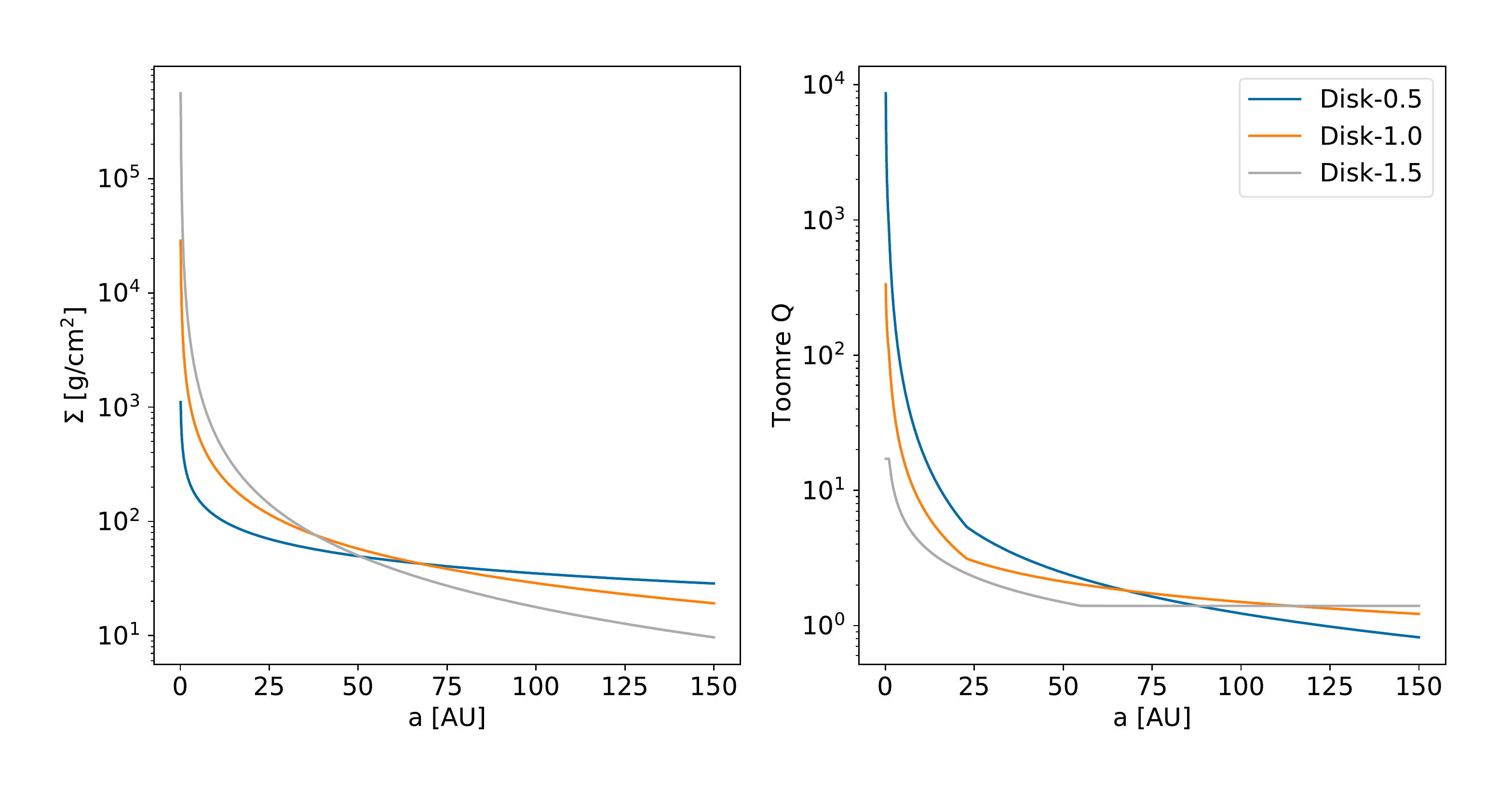}
	\caption{Surface density (left) and Toomre parameter (right) as a function of orbital radius.}
	\label{fig:disk_profiles}
\end{figure}

We assume that clumps form in a Toomre unstable disk with $Q_{min} < 1.4$ \citep{Durisen2006} but after fragmentation, the disk is taken to be static, i.e., its dynamical evolution is not followed.  We then consider three disk models with commonly used power-law profiles for the surface density $\Sigma \propto a^{-\sigma}$ as a function of orbital radius $a$. The surface density and Toomre profiles of the disk models are shown in \cref{fig:disk_profiles} and their properties are summarized in \cref{tab:disk_profiles}. 
For model Disk-1.0 we also vary the stellar masses while keeping the ratio of disk-to-star mass constant. 
The temperature profiles are derived from:

\begin{equation}
T = \text{max}\big[\text{min}\left(T_{0}(\frac{a}{\textup{AU}})^{-5/4}, \, T_{0}\right), \, 30 \, \text{K}\big],
\end{equation}

where $T_{0} = 1500 \text{K}$. For the third disk model, this leads to a Toomre profile that is too high for disk fragmentation and therefore in this case we set the temperature in the outer region to 15 K so the disk is gravitationally unstable.

The total disk mass is taken to be 30 \% of the stellar mass and the Shakura-Sunyaev $\alpha$ parameter is set to $\alpha = 0.005$, which is based on results of turbulence generation studies \citep{Nelson2004}.  
In this work we use $\alpha = \alpha(\mu, Q)$ as derived in \cite{Kratter2007}, where $\mu$ is the disk-to-total mass ratio given by $\mu = M_{\text{disk}}/\left(M_{\text{disk}} + M_{\text{star}}\right)$  and $Q$ is the \textit{minimal} Toomre parameter. Although the disk's viscosity is expected to slightly change for different disk-to-star mass ratios \citep{Vorobyov2009}, for simplicity we keep a constant $\alpha$ also when we consider different disk and stellar masses. 
\par

\begin{table}[h]
	\centering
	\begin{tabular}{cccc}
		\hline 
		 & Disk-0.5 & Disk-1.0 & Disk-1.5 \\ 
		\hline 
		\rule[-1ex]{0pt}{2.5ex}	Stellar mass [M$_{\astrosun}$] & 1.0 & 1.0 & 1.0 \\ 
		\rule[-1ex]{0pt}{2.5ex} Disk mass [M$_{\astrosun}$] & 0.3 & 0.3  & 0.3  \\ 
		\rule[-1ex]{0pt}{2.5ex} Surface density [$\text{g}/\text{cm}^{-2}$] & $21.0 \, (a/(100 \text{AU}))^{-1/2}$ & $17.3 \, (a/(100 \text{AU}))^{-1}$ & $14.4 \, (a/(100 \text{AU}))^{-3/2}$ \\ 
		\hline 
	\end{tabular}
	\caption{Disk profiles with the different surface density power-laws. Disk-$\sigma$, where $\sigma =$ 0.5, 1.0, 1.5 is the exponent in $\Sigma \propto a^{-\sigma}$. The densities are scaled to ensure a constant disk-to-star mass ratio.}
	\label{tab:disk_profiles}
\end{table}

\subsection{Initial conditions} \label{sec:initial_conditions}
The primordial population of clumps is generated using two different sets of initial conditions. 
We assume that clumps form beyond 80 AU where cooling is efficient. However, the topic of disk cooling is still being debated; recent models using the $\beta$ cooling approximation suggest that disks could fragment at small radial distances, even within 20 AU \citep{Boss2017}. For both sets, the initial orbital radius $a$ is randomly chosen in the range of $80 - 120$ AU, where the disk is likely to fragment. The first set of initial conditions (hereafter ICL) includes clumps with radii of $R = 1.0 - 6.0$ AU and masses of $M_{\text{Clump}} = 0.5 - 5.0 M_{\text{J}}$. The initial radius and mass of a clump is randomly chosen from the given intervals. The second set of initial conditions (ICLM) includes clumps with radii in the range $R = 2 - 12 $ AU and masses between 5 and 12 $M_{\text{J}}$. The latter set could reflect a scenario of small clumps mergers. The initial clump mass and radius are chosen randomly in these intervals, as discussed below we find that the results are insensitive to the initial radius, as the memory of the initial size is rapidly determined by contraction and tidal downsizing. In order to investigate the sensitivity of the final population to the initial clump mass, we generate additional sets of initial conditions by either subtracting (IC-Lighter) or adding (IC-Heaver) $4 \, M_{\text{J}}$ to the initial conditions ICL and ICLM.

\subsection{Migration and clump contraction} \label{sec:migration_and_clump_contraction}
The survival or destruction of clumps is the result of a competition between processes that either lead to contraction or rip it apart. Essentially, these are the clump's contraction due to self-gravity and the tidal disruption of the host star. The clump's survivability also depends on how quickly the clump can accrete mass, migrate inwards, and reach the second core collapse (dynamical collapse). The contraction (pre-collapse) timescale is set to the time required for dissociation of molecular hydrogen in the clump's center. Following 3D simulations, the collapse is assumed to occur when the dissociation parameter, i.e., the ratio between protons in the atomic hydrogen and the total number of protons in the gas, is 1\%. Typical collapse timescales are found to be $\sim 5000-6000$ years \citep{Galvagni2012}. 

We also consider another option in which, depending on the clump's mass, we assign it a certain collapse timescale based on the results of \citet{Helled2009}. This determines the pre-collapse evolution timescale which ranges between $10^{3}$ and $10^{5}$ years instead of a few $10^3$ years. In both cases, the time to reach dynamical collapse is significantly shorter than the typical disk lifetime of a few $10^{6}$ years \citep{Mamajek2009, Hillenbrand2005}. 
It should be noted, however, that self-gravitating disks are expected to evolve relatively fast ($< 10^6$ years) 
due to outward angular momentum transport via gravitational torques induced by spiral density waves \citep{Durisen2006}. If this is the case, evolving clumps for several $10^5$ years could be comparable to the disk's lifetime. 

After clumps form they can migrate inward due to their interactions with the disk.
The migration timescale follows the following description \citep{Baruteau2011}: 
\begin{equation}
\label{eq:migration}
\tau = P_{\text{orb}} \, 5.6 \, (3.8 - \sigma)^{-1} \gamma Q\left(\frac{q}{(h/a)^{3}}\right)^{-1}\left(\frac{h/a}{0.1}\right)^{-2}
\end{equation}
where $P_{\text{orb}}$ is the orbital period, $\gamma = 5/3$ is the adiabatic index, $q$ the ratio between clump and stellar mass, $\sigma$ the exponent of the disk's surface density power-law, and $h$ is the disk scale height in vertical hydrostatic equilibrium. For all the cases we consider the migration timescales are at most $\sim10^{4}$ years, which are comparable to the contraction timescales. 

\subsection{Tidal downsizing} \label{sec:tidal_downsizing}
As a clump migrates inward, its outermost edges are affected by stellar tidal forces. If the clump migrates fast enough and is still extended, its outer layer can be stripped away. This occurs when the clump's  radius is comparable to its Hill radius, i.e., when $R_{\text{clump}} = R_{\text{H}} = a \left(\frac{M_{\text{clump}}}{3 \, M_{\text{star}}}\right)^{1/3}$. We include tidal downsizing when $R_{\text{clump}} = 1/3 \, R_{\text{H}} $ as derived from 3D simulations \citep{Galvagni2012}. Tidal downsizing implies mass loss, which in turn leads to an increase in the migration timescale. Then the clump has a chance to contract further and survive tidal disruption. 

We consider clumps as being tidally destroyed if:
\begin{itemize}
	\item A clump's mass or radius drops below $10^{-5}$ of a Jupiter mass/radius.
	\item The orbital radius is smaller than 0.05 AU and its migration timescale is shorter than a few years.
\end{itemize}

Although the second condition is mostly chosen for numerical stability, it also has a physical meaning: at this point, the protoplanet would be heavily affected by the stellar magnetic field and tides while migrating very quickly and is expected to be   destroyed.

\subsection{Clump density profiles} \label{sec:clump_density_profiles}
In order to investigate the influence of the clump's internal structure on its evolution, we consider clumps with varying density profiles (spherical). The first is a quasi-homogeneous profile where the tidal mass loss is given by a semi-analytical formula that depends on a simple mass-scaling \citep{Galvagni2014}. 
The second and third structures are a simple homogeneous and a point-like density profiles. Finally, we also consider a more realistic power-law density profile where the density is given by  $\rho(r) = Ar^{\beta} + B$, where $A$ and $B$ are constants determined by boundary conditions. 
In both of these structures, which are clearly more realistic, the outermost density is taken to be ten times the local disk surface density.

\subsection{Mass accretion} \label{sec:mass_accretion}
As clumps migrate through the gaseous disk they can also accrete mass (gas) from the disk. The mass accretion rate is given by \citep{Galvagni2014}:

\begin{equation}
\frac{dM}{dt} = 1 \times 10^{-7} \times 3^{\log(\Sigma/\Sigma_{100})}\left(\frac{M_{\text{clump}}}{M_{\text{J}}}\right)^{2/3}\left(\frac{M_{\text{star}}}{M_{\astrosun}},\right)^{-1/6} \, M_{\astrosun} \text{yr}^{-1}
\end{equation}

where $\Sigma_{100} = \Sigma(a = 100 \, \text{AU})$ is the surface density at 100 AU. 
In our baseline case, we assume that once a clump opens a gap the local surface density is small and therefore mass accretion terminates. While the clump is accreting mass, in order to contract further it must radiate energy away. We therefore limit the mass accretion rate so it cannot exceed the Kelvin-Helmholtz timescale. As the clump grows in mass we scale the radius accordingly by the following scaling: $r'= r + \delta r = r \left(1 + \frac{\delta M}{M_{\text{clump}}}\right)$. 

\subsection{Gap opening} \label{sec:gap_opening}
Considering the exchange of angular momentum between the clump and the gas disk is important for massive objects. 
When a massive clump migrates within the disk, it can open a gap if its gravitational torque is large enough to overcome the local viscous and pressure torque. The exact criterion for gap opening is in fact unknown, and therefore we explore three different gap opening criteria. The first is the standard torque balance criterion \citep{Crida2006, Kley2012}:

\begin{equation}
\label{eq:gapopening}
\mathcal{P} = \frac{3}{4} \frac{h(a)}{R_{\text{H}}} + \frac{50}{q\mathcal{R}} \leq 1,
\end{equation}

where $\Omega$ is the Keplerian orbital frequency. The Reynolds number is given by  $\mathcal{R} = a^{2} \Omega^{2}/\nu = a^{2} \Omega/\left(\alpha h^{2}\right)$. Whenever a gap is opened, the migration timescale becomes comparable to the viscous diffusion timescale $\tau_{\text{visc}} = a^{2}/\nu$ \citep{Lin1986a}, which is the time it takes for viscosity to smooth out surface density gradients and is typically between $10^{5}$ and $10^{6}$ years. 

A modified version of this criterion can be one which includes a comparison between the viscous and crossing timescales $\tau_{\text{cross}} = R_{\text{HS}}\left(da/dt\right)^{-1}$ with $da/dt$ being the migration rate \citep{Malik2015}. 
$\tau_{\text{cross}}$ is the timescale it takes a clump to pass/cross over a typical gap size while $R_{\text{HS}}$ is the distance traveled before a gap opens again, and is taken to be $2.5 R_{\text{H}}$ following \citet{Paardekooper2009}. In addition to the standard gap opening condition, the second condition includes $\tau_{\text{visc}} < \tau_{\text{cross}}$.

Alternatively, the additional criterion can use the characteristic gap formation timescale \citep{Lin1986a}: 
$\tau_{\text{gap}} = \left(M_{\text{star}}/M_{\text{clump}}\right)^{2} \left(h/a\right)^{5} \Omega^{-1}$.
Hydrodynamical simulations of massive planets migrating in self-gravitating disks suggest that the timescale inferred from the analytical estimate for gap opening  is too short \citep{Malik2015}. Thus, a correction factor $\eta \sim 100-1000$ must be applied to align the analytical expression with numerical results. Instead of using the viscous timescale, the third gap opening criterion becomes $\eta \, \tau_{\text{gap}} < \tau_{\text{crossing}}$. As a result, the clump opens a gap only if the gap opening timescale is shorter than the time it takes to cross over the gap region. To summarize, the following gap opening criteria are considered:

\begin{enumerate}
	\item when $\frac{3}{4} \frac{h(a)}{R_{\text{H}}} + \frac{50}{q\mathcal{R}} \leq 1$ is satisfied
	\item if in addition to the first condition $\eta \, \tau_{\text{gap}} < \tau_{\text{cross}}$
	\item if in addition to the first condition $\tau_{\text{visc}} < \tau_{\text{cross}}$
\end{enumerate}

\section{Results} \label{sec:results}
For every simulation and model assumption we generate 500 clumps of each initial condition (ICL and ICLM) and evolve them separately including clump evolution, migration, and tidal disruption. Since we want to explore the case when clumps open gaps, we use $\alpha = 0.005$. The evolution is followed until the clump reaches second core collapse or the clump is tidally destroyed. The explored combinations of gap opening and disk profiles are labeled and summarized in \cref{tab:combinations}. In the following sections, tidally destroyed clumps are always excluded when we present the physical properties of the inferred population.

\begin{table}[h]
	\centering
	\begin{tabular}{lll}
		\hline
		Label & Gap opening criterion & Disk profile \\ \hline
		Disk-1.0-Crida & $\frac{3}{4} \frac{h(a)}{R_{\text{H}}} + \frac{50}{q\mathcal{R}} \leq 1$ & $\sigma = 1.0$ \\
		Disk-1.0-Gap & $\eta \, \tau_{\text{gap}} < \tau_{\text{cross}}$ 					  & $\sigma = 1.0$ \\
		Disk-1.0-Viscous & $\tau_{\text{visc}} < \tau_{\text{cross}}$                       & $\sigma = 1.0$ \\ 
		Disk-$\sigma$ & $\eta \, \tau_{\text{gap}} < \tau_{\text{cross}}$ & $\sigma = 0.5, 1.0, 1.5$ \\ \hline
	\end{tabular}
	\caption{The different model combinations we consider. For the new gap opening criteria $\eta \tau_{\text{gap}} < \tau_{\text{cross}}$ and $\tau_{\text{visc}} < \tau_{\text{cross}}$, the clump must implicitly satisfy the torque balance criterion given in addition to the timescale criterion. The disk surface density scales as $\Sigma \propto a^{-\sigma}$.}
	\label{tab:combinations}
\end{table}

\subsection{Gap opening comparison} \label{sec:gap_opening_comparison}
\begin{figure}[h]
	\centering
	\includegraphics[scale=0.35]{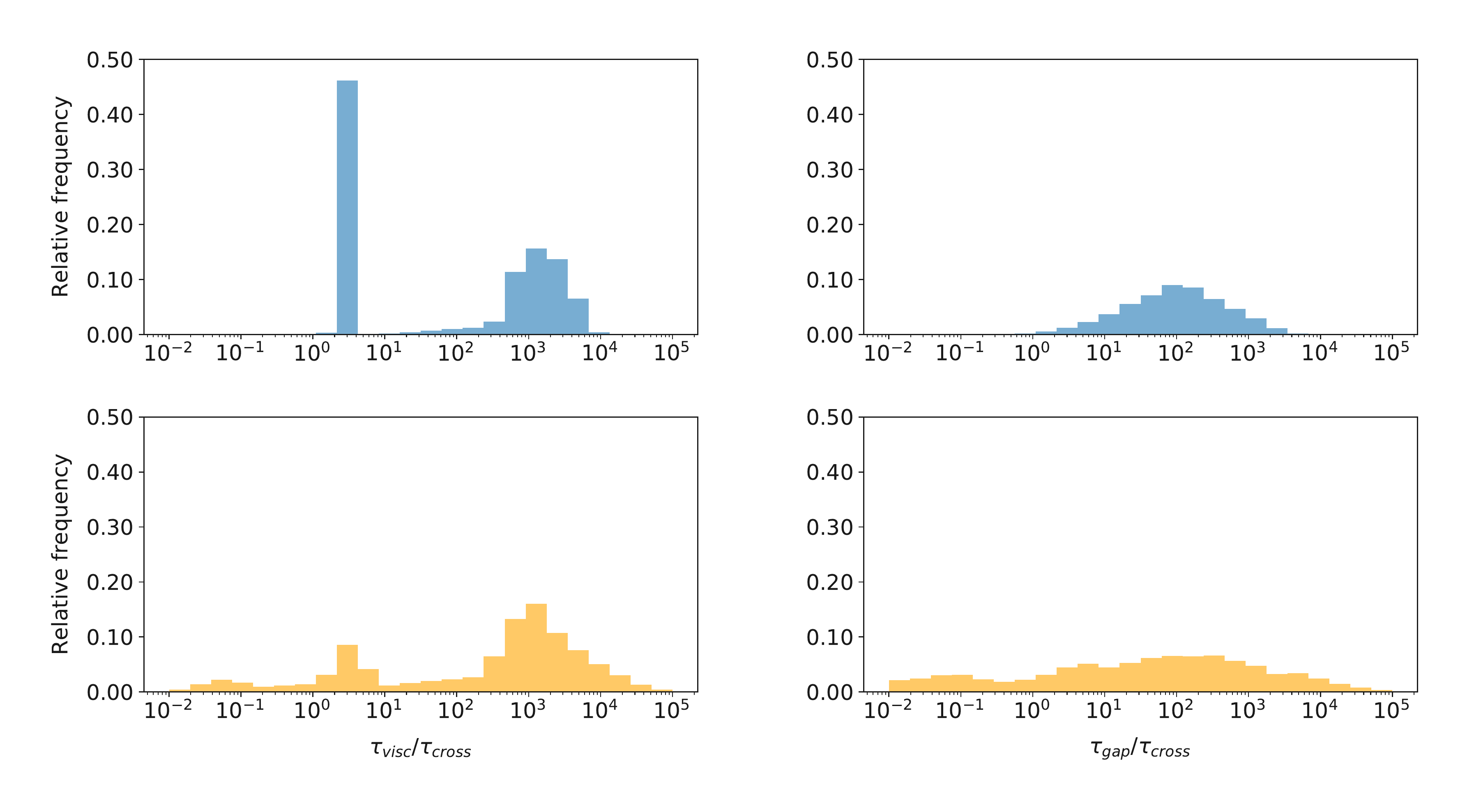}
	\caption{A timescale comparison for the models Disk-1.0-Crida (top) and Disk-1.0-Gap (bottom). The peak in the top left plot is linked to massive clumps that open gaps and then migrate on a viscous timescale.}
	\label{fig:timescales_fraction}
\end{figure}

We first investigate how many clumps open a gap when using the different gap opening criteria. 
In this section we use Disk-1.0 and  the power-law clump density profile. We find that in the models Disk-1.0-Gap/Viscous with $\eta = 1000$ gaps are nearly never opened (see \cref{tab:gapcounts} for a direct comparison). The gap and viscous timescale criteria lead to a similar gap opening frequency. Figure \ref{fig:timescales_fraction} compares the different timescales at every time-step during the clump's evolution. For a gap to be opened, the ratio in both cases should be below one. For Disk-1.0-Crida this never happens. The viscous and gap opening timescale are generally orders of magnitude larger than the crossing time, with the viscous timescale typically being the largest of the three. In a few cases, the ratio is low enough, but still no gaps are opened. This implies that it is the torque balance criterion that is not satisfied. As suggested by \cref{fig:timescales_fraction}, the ratio $\tau_{\text{gap}}/\tau_{\text{cross}}$ is usually smaller than  $\tau_{\text{visc}}/\tau_{\text{cross}}$. The correction factor for the analytical gap opening timescale can change these results. Hereafter we use $\eta = 100$ as our baseline.

\begin{table}[h]
	\begin{tabular}{lc}
		\hline
		Model & Gap frequency  \\ \hline
		Disk-1.0-Crida: ICL  & 0.25        \\
		Disk-1.0-Crida: ICLM & 1.00       	\\
		Disk-1.0-Gap/Viscous: ICL 	 & 0.01		\\
		Disk-1.0-Gap/Viscous: ICLM 	 & 0.00		\\ \hline							
	\end{tabular}
	\centering
	\caption{Gap opening probabilities for the different gap opening criteria. ICL and ICLM correspond to the light and heavy initial mass function, where $M_{\text{clump}}$ is between $0.5$ and $5.0 \, M_{\text{J}}$ and $5 - 12 \, M_{\text{J}}$, respectively.}
	\label{tab:gapcounts}
\end{table}

\subsection{Physical properties of clumps for different gap opening criteria} \label{sec:clump_property_distributions}
In this section we compare the inferred physical properties of clumps for the different gap opening criteria. The clumps' physical properties presented in the histograms are taken at the point when the clump has reached dynamical collapse. As mentioned above, we exclude clumps that are tidally destroyed in all histograms. For this comparison, we use the models Disk-1.0-Crida/Gap. We find that the clump's survival probability is strongly influenced by gap opening. If we only consider the torque balance criterion, the survival rate is 0.67, while for the criterion using the gap opening timescale, 12\% of the clumps survive. 

More clumps survive in general when gaps open - this is especially true for the more massive clumps as they migrate quickly inwards where they are destroyed by stellar tidal stripping. When gaps open, even the massive clumps can survive, since from that point on they evolve on a viscous timescale. It can be seen from \cref{tab:gapcounts} that heavy clumps open gaps more frequently than lighter ones, which leads to higher survival rates. 

\begin{figure}[h]
	\centering
	\includegraphics[scale=0.35]{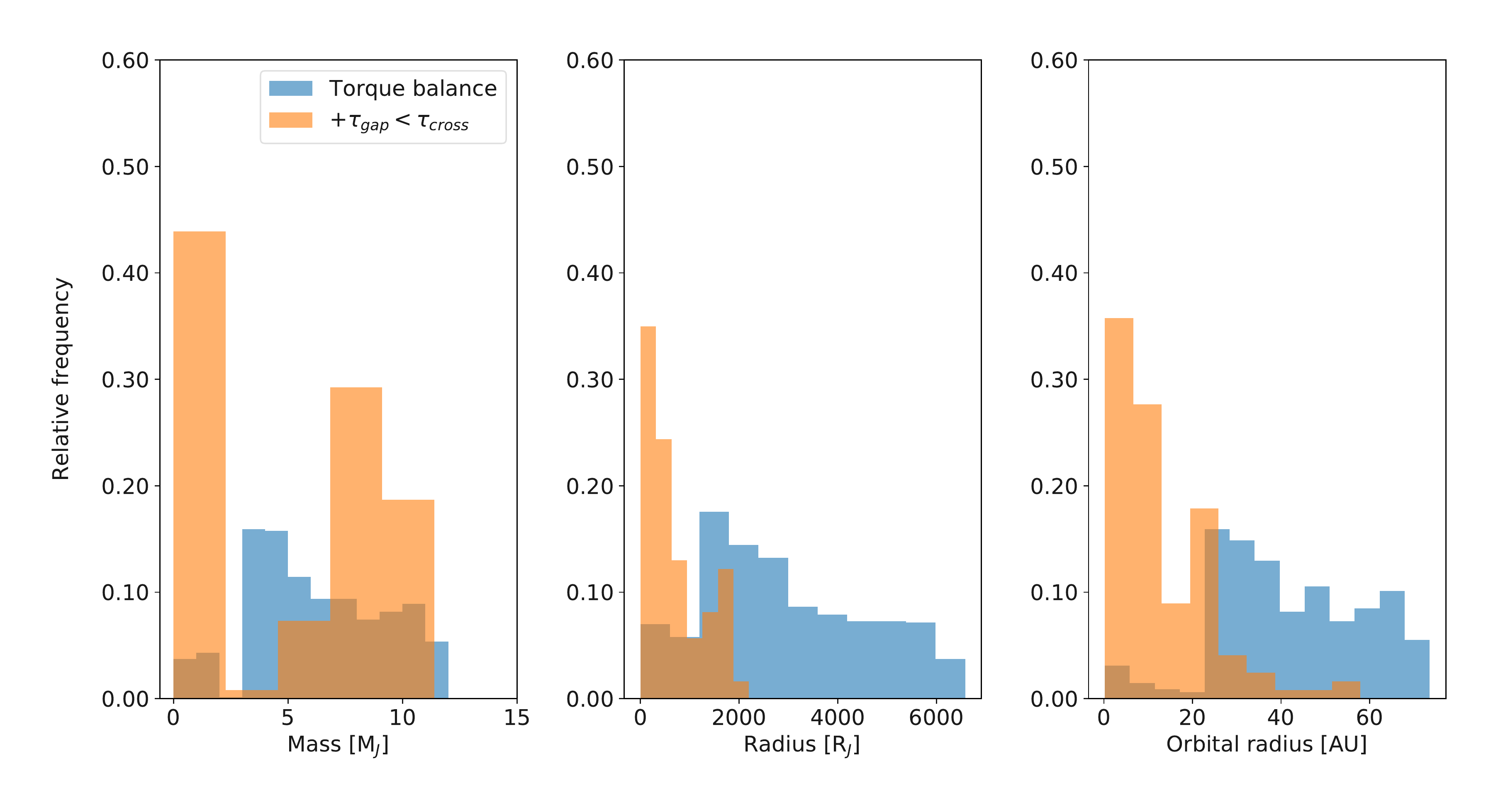}
	\caption{Inferred mass, radius, and orbital radius of the surviving clumps for Disk-1.0-Crida (blue) and Disk-1.0-Gap (orange). 
	The bins are normalized to show relative frequencies. The new gap opening criterion leads to a significant shift towards smaller clumps orbiting closer to the central star in Disk-1.0-Gap (see text for details).}
	\label{fig:oldvsvisc_oldprofile}
\end{figure}

Figure \ref{fig:oldvsvisc_oldprofile} shows the distribution of clump mass, radius, and orbital radius of all the surviving clumps for the different modeled combinations. The clumps final masses are predominantly in the range of $0.1 - 12 M_{\text{J}}$. The clumps' radius distribution shows that all the surviving clumps evolved significantly, with most of the radii being  in the range of $1 - 10^{3} R_{\text{J}}$. The final distribution of radial distances suggests that surviving clumps can have both short and long orbital periods with the range being $10^{-1} - 10^{2}$ AU. Typically, lighter clumps that do not open gaps end up farther out due to their slower migration in comparison to heavier clumps. 
 
Our results clearly demonstrate that the possibility of gap opening alters both the survival probability and the inferred population. Clumps that open gaps have survival probabilities several times higher than those that do not open gaps, and are also larger in size, more massive, and orbit farther out. Faster migration decreases the clumps' survivability due to tidal downsizing, which is especially detrimental to heavy clumps. We find that when we consider the gap opening timescales we get a rearrangement in the surviving population towards smaller clumps orbiting closer to the star.

\subsection{Gap opening timescale revisited} \label{sec:gap_opening_time_scale_revisted}
We introduced a scaling factor for the gap opening timescale $\eta = 100$ to match the analytical formula to simulations results. It is unclear which $\eta$ value is most appropriate. While here we use $\eta = 100$ as the baseline case, this value depends on the model and/or computational routine, and therefore should not be taken as an optimized-value. To investigate the influence of the $\eta$ value on the results, we also consider $\eta = 1000,100,10,1$ and check gap opening probability. Again, 500 clumps are simulated for each initial condition. For the initial condition of massive clumps, we find that clumps rarely open gaps when $\eta=100$. For $\eta= 10$ the gap opening probabilities start to match those found in \cref{tab:gapcounts} in the case where only the first gap opening criterion is considered. Also the survival probabilities match those found for the conditions Disk-1.0-Crida when $\eta = 10$. Setting the scale factor even lower does not change the results significantly. The results are summarized in \cref{tab:eta}. It should be noted that empirical results suggest that small $\eta$ values are less realistic, which in turn means that gaps are rarely opened. In any case, we find that typically $\tau_{\text{gap}} \gg \tau_{\text{cross}}$ (with the correction factor applied), which is in agreement with previous results \citep{Malik2015}. This shows that migration is sensitive towards a change in the scaling factor $\eta$.

\begin{table}[h]
	\centering
	\begin{tabular}{lccc}
		\hline
		& ICL & ICLM & Survival rate  \\ \hline
		$\eta = 1000$ & 0.01    & 0.00	& 0.05      \\
		$\eta = 100$  & 0.01    & 0.14	& 0.12      \\
		$\eta = 10$   & 0.25    & 1.00	& 0.67      \\
		$\eta = 1$    & 0.25    & 1.00	& 0.67      \\ \hline
	\end{tabular}
	\caption{Gap opening and survival probabilities for different scaling factors $\eta$ in the model Disk-1.0-Gap. When the scaling factor $\eta$ is reduced to 10, the probabilities match those of Disk-1.0-Crida.}
	\label{tab:eta}
\end{table}

\subsection{Properties of surviving clumps} \label{sec:properties_of_surviving_clumps}
We next explore the distribution of the masses and radial distances of the surviving clumps, and divide their populations as follows: closer (or farther) than 10 AU and, above (or below) 10 $M_{\text{J}}$. We compare the results of a model with an efficient gap opening (Disk-1.0-Gap with $\eta = 10$) and one with  $\eta = 100$ (rare gap opening). 

When $\eta = 100$ clumps with masses above 10 $M_{\text{J}}$ rarely survive, while clumps with lower masses are equally distributed inside and outside $10$ AU. For $\eta = 10$, 14\% of the surviving clumps heavier than 10 $M_{\text{J}}$ orbit outside of 10 AU. Of the lighter clumps, 81\% orbit farther out than 10 AU, demonstrating that efficient gap opening inserts a bias towards larger orbital radii. The results are summarized in \cref{tab:counts2}. \\

\clearpage
\begin{table}[h]
	\centering
	\begin{tabular}{lll}
		\hline
		& $\eta = 100$ & $\eta = 10$ \\ \hline
		a \textless 10 AU \& M \textless 10 $M_{\text{J}}$:      & 0.52         & 0.04        \\
		a \textgreater 10 AU \& M \textless 10 $M_{\text{J}}$:    & 0.37         & 0.81        \\
		a \textless 10 AU \& M \textgreater 10 $M_{\text{J}}$;    & 0.00         & 0.00        \\
		a \textgreater 10 AU \& M \textgreater 10 $M_{\text{J}}$: & 0.11         & 0.14     	\\ \hline  
	\end{tabular}
	\caption{Relative frequency of clumps orbiting inside (outside) a radius of 10 AU split into two groups based on their mass. The numbers reflect the sensitivity of the inferred population on the scaling factor $\eta$.}
	\label{tab:counts2}
\end{table}

\subsection{Clump density profiles} \label{sec:density_profiles}
\begin{figure}[h]
	\centering
	\includegraphics[scale=0.32]{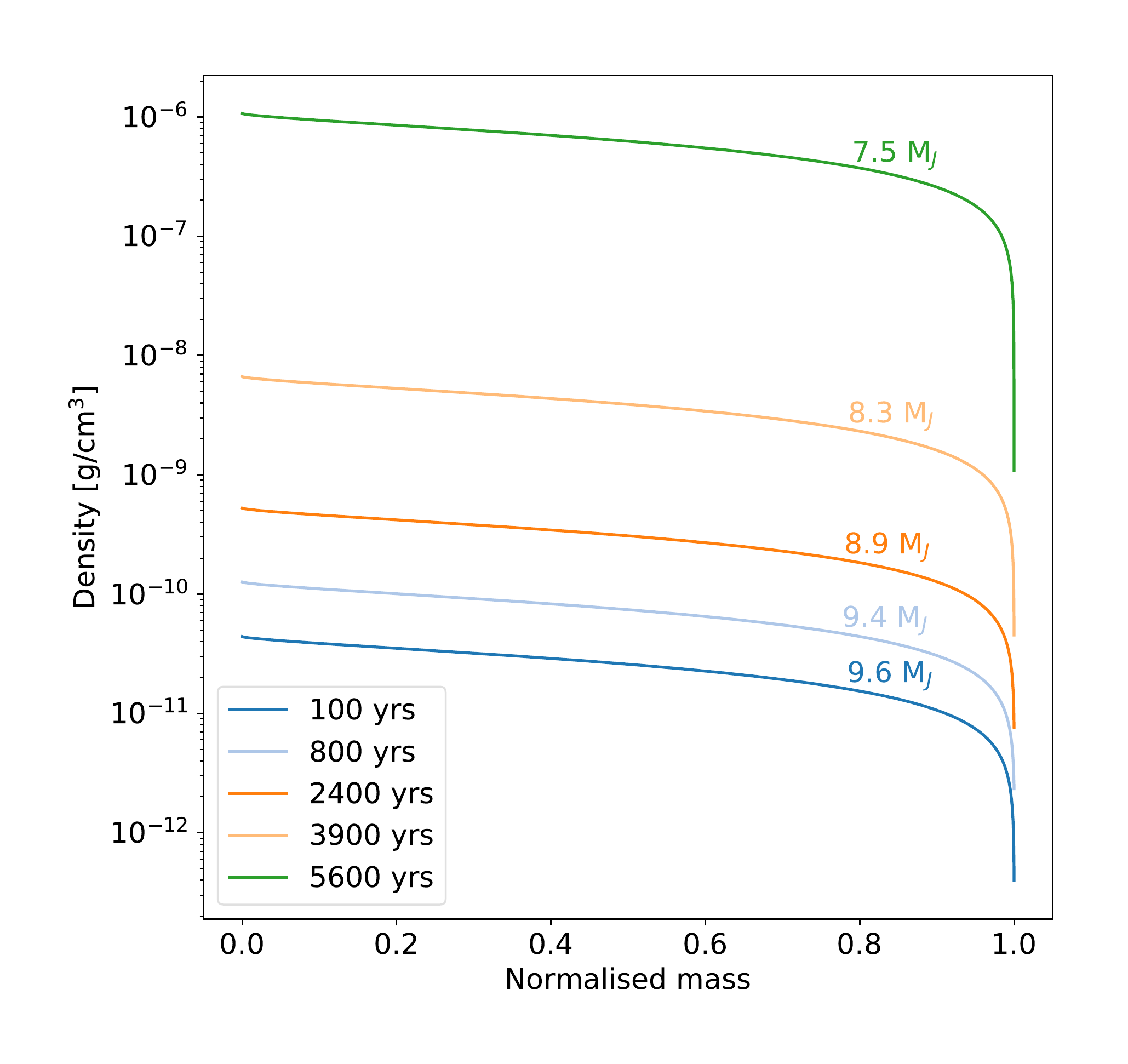}
	\caption{Clump density profiles vs.~normalized mass at different times. The lines are labeled according to the clump's current mass. In order to survive tidal downsizing, clumps must  contract and become significantly denser (see text for discussion).}
	\label{fig:density_evolution}
\end{figure}

We next investigate the influence of the clump density profiles on the resulting population. Here we use disk model Disk-1.0-Gap with $\eta = 100$. The evolution of the density profile for the power-law profile is shown in \cref{fig:density_evolution}. As can be seen from the figure, the clump becomes denser by a few orders of magnitude as time progresses as expected from detailed evolution calculations \citep{Helled2010, Vazan2012} and high resolution studies of clumps \citep{Galvagni2012, Szulagyi2016}.

\begin{figure}
	\centering
	\begin{tabular}{ccc}
		\includegraphics[width=56mm]{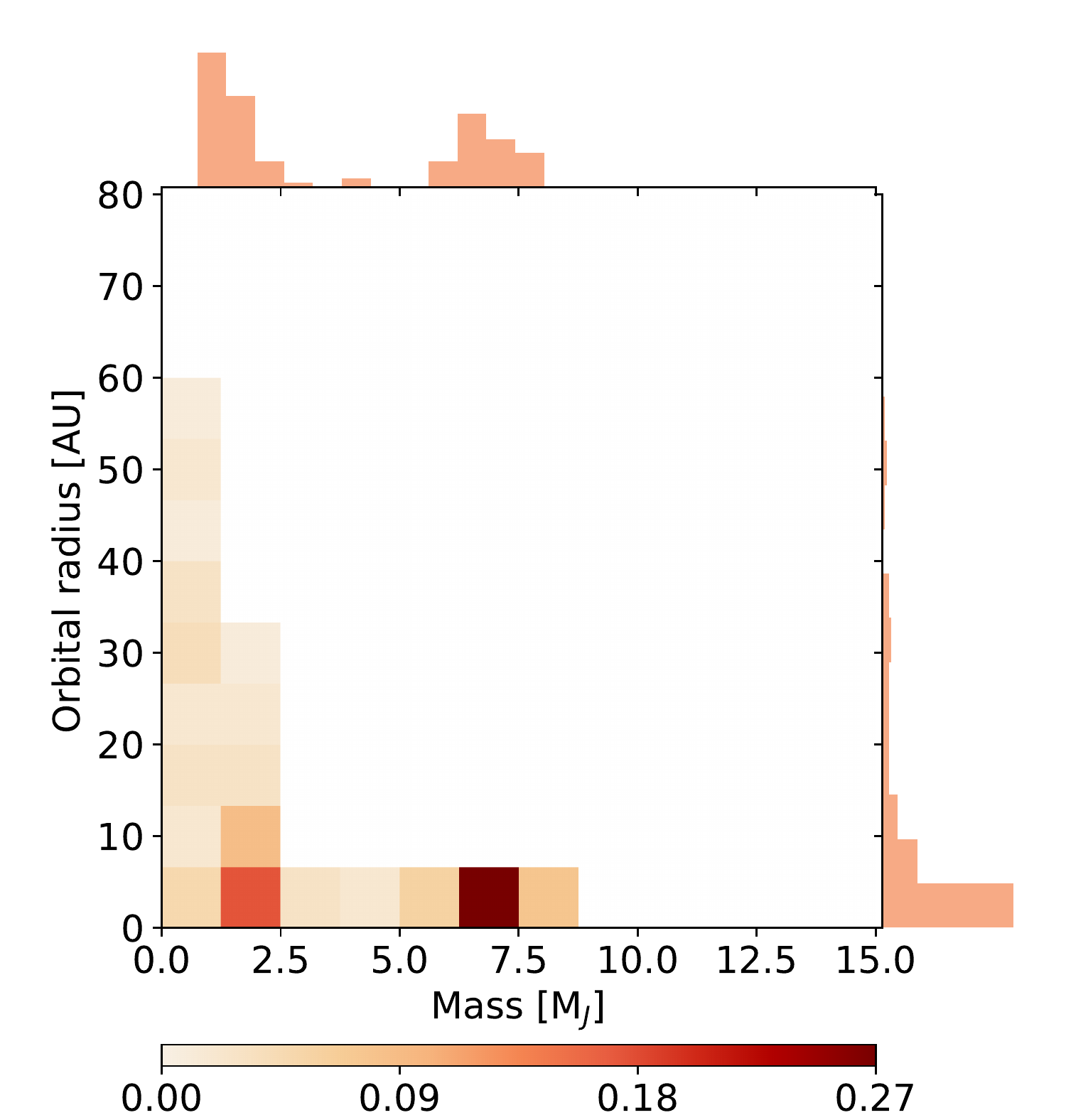} &   \includegraphics[width=56mm]{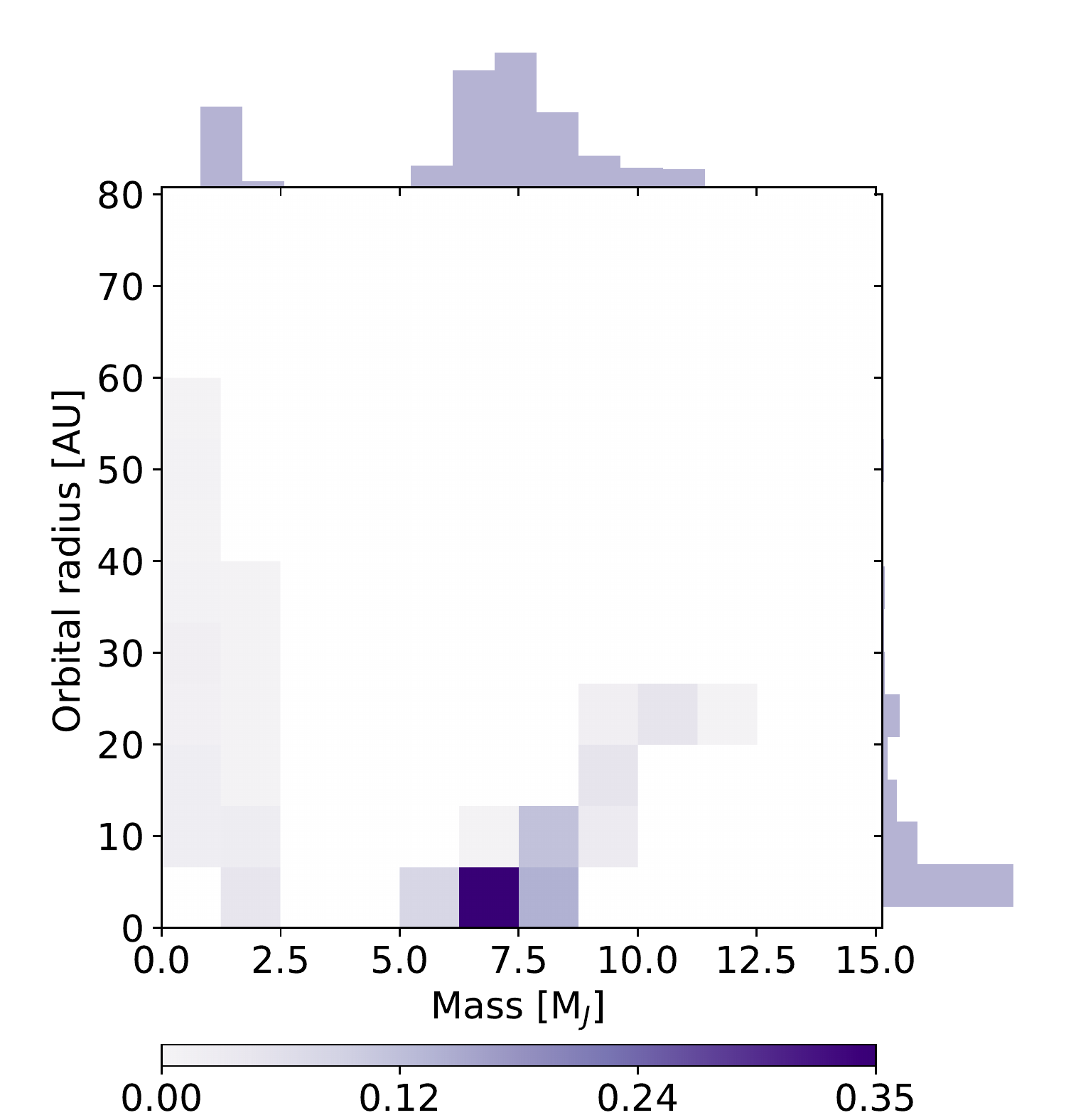} & \includegraphics[width=56mm]{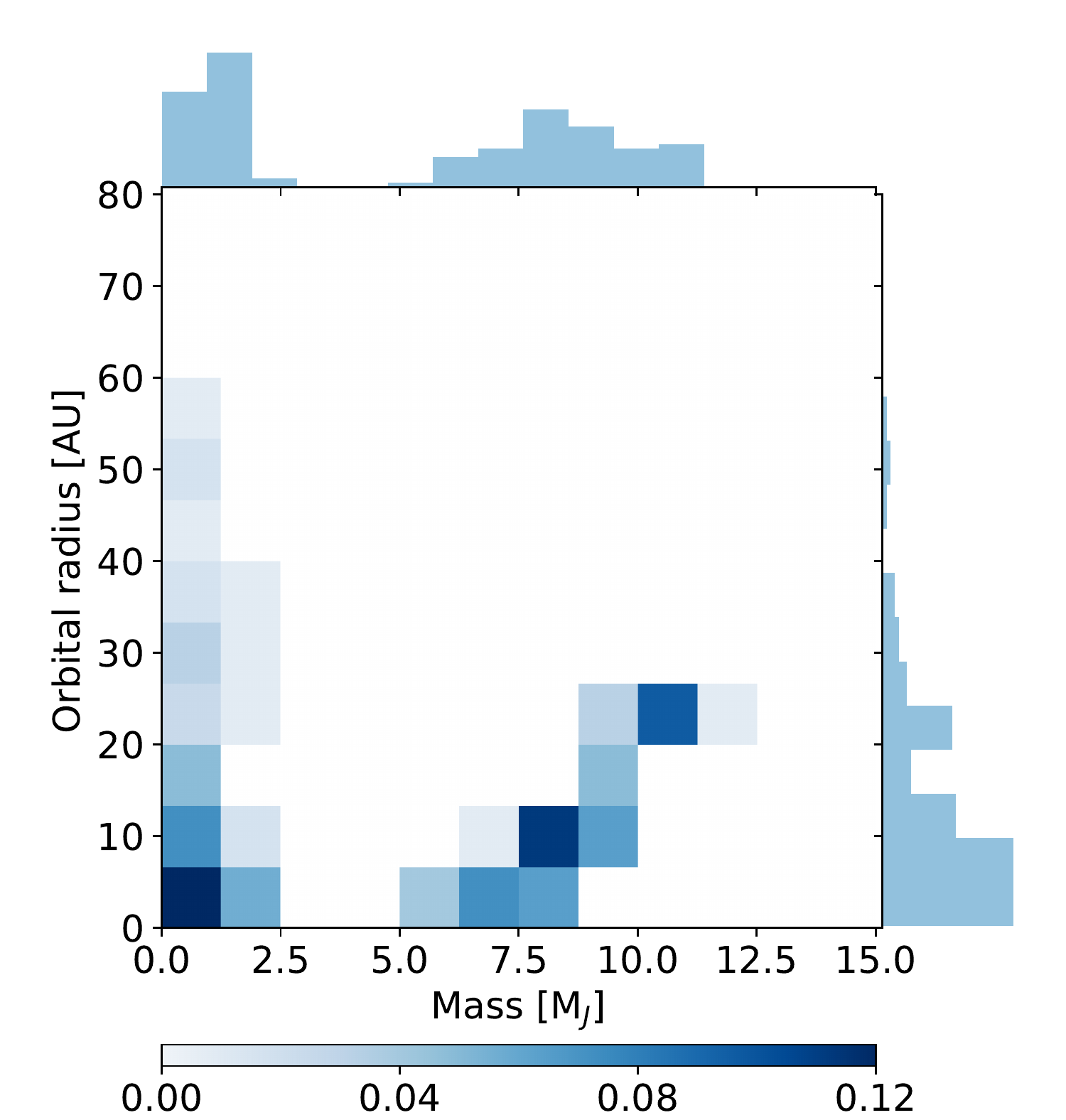} \\
		(a) Quasi-homogeneous & (b) Point-like  & (c) Power-law \\[6pt]
	\end{tabular}
	\caption{Histograms of mass and orbital radius of the surviving population for the different clump density profiles. The colorbar shows the relative frequency. }
	\label{fig:density_comparison}
\end{figure}

The resulting clump population is shown in \cref{fig:density_comparison}, while \cref{tab:density_profiles} summarizes the key properties. The clump's density profile clearly has a crucial influence on the inferred population, and therefore must be modeled correctly in population synthesis models.  In the point-like and the realistic power-law density profiles, the outer regions of the clump carry little mass and the clumps are less affected by tidal downsizing. This leads to a population of massive clumps at larger orbits (20 - 30 AU) in comparison to the quasi-homogeneous case. The complex dynamics between tidal downsizing and gap opening is largely determined by the clump's structure. In contrast to the rare gap opening in the quasi-homogeneous density model, clumps with realistic density profiles open gaps more frequently, and thus the surviving clumps orbit further from the star.

\begin{table}[h]
	\centering
	\begin{tabular}{lcccc}
		\hline 
		\rule[-1ex]{0pt}{2.5ex} Density profile & Gap opening & Survival & $a$ \textless 10 AU & $M$ \textgreater 10 M$_{\text{J}}$ \\ 
		\hline 
		\rule[-1ex]{0pt}{2.5ex} Quasi-homogeneous & 0.04 & 0.10  & 0.75  & 0.00 \\  
		\rule[-1ex]{0pt}{2.5ex} Point-like & 0.21  & 0.26 & 0.75 & 0.05  \\ 
		\rule[-1ex]{0pt}{2.5ex} Power-law & 0.08 & 0.12 & 0.52  & 0.11 \\ 
		\hline 
	\end{tabular}  
	\caption{The sensitivity of the results to the clump's density profile. Gap opening and survival probabilities (first two columns) and frequency of clumps that orbit within 10 AU or are heavier than 10 $M_{\text{J}}$ (last two columns).}
	\label{tab:density_profiles}
\end{table}

None of the clumps with homogeneous densities survive. The quasi-homogeneous profile is unrealistic and leads to much smaller masses, and a lack of surviving clumps with masses larger than 8 $M_{\text{J}}$. All the heavy clumps that survive orbit close to the star. The inferred populations of the point-like and power-law profiles are similar qualitatively, but a closer look at the frequencies in \cref{tab:density_profiles} reveals that there are differences. The point-like case leads to a rather sharp peak of clumps orbiting below 10 AU with masses of $\sim$1 $M_{\text{J}}$. Using a power-law density profile leads to a distribution that is more continuous in both orbital radius and mass. In all cases, we find no massive clumps orbiting at large radial distances. Hereafter, we use the power-law density profile as the baseline case. \\

\subsection{Clump initial mass function} \label{sec:mass_function}
\begin{figure}
	\centering
	\begin{tabular}{ccc}
	\includegraphics[width=56mm]{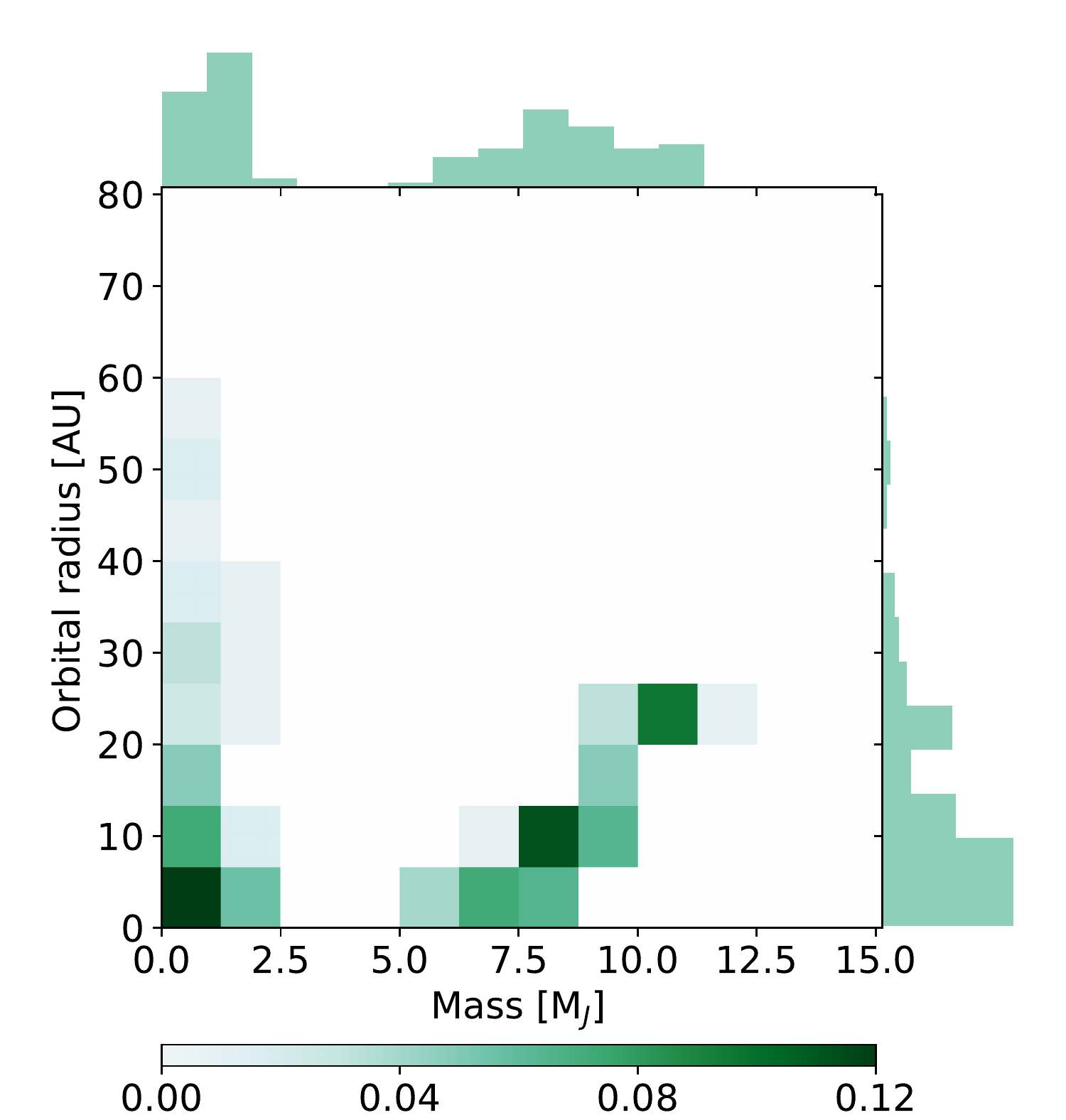} &   \includegraphics[width=56mm]{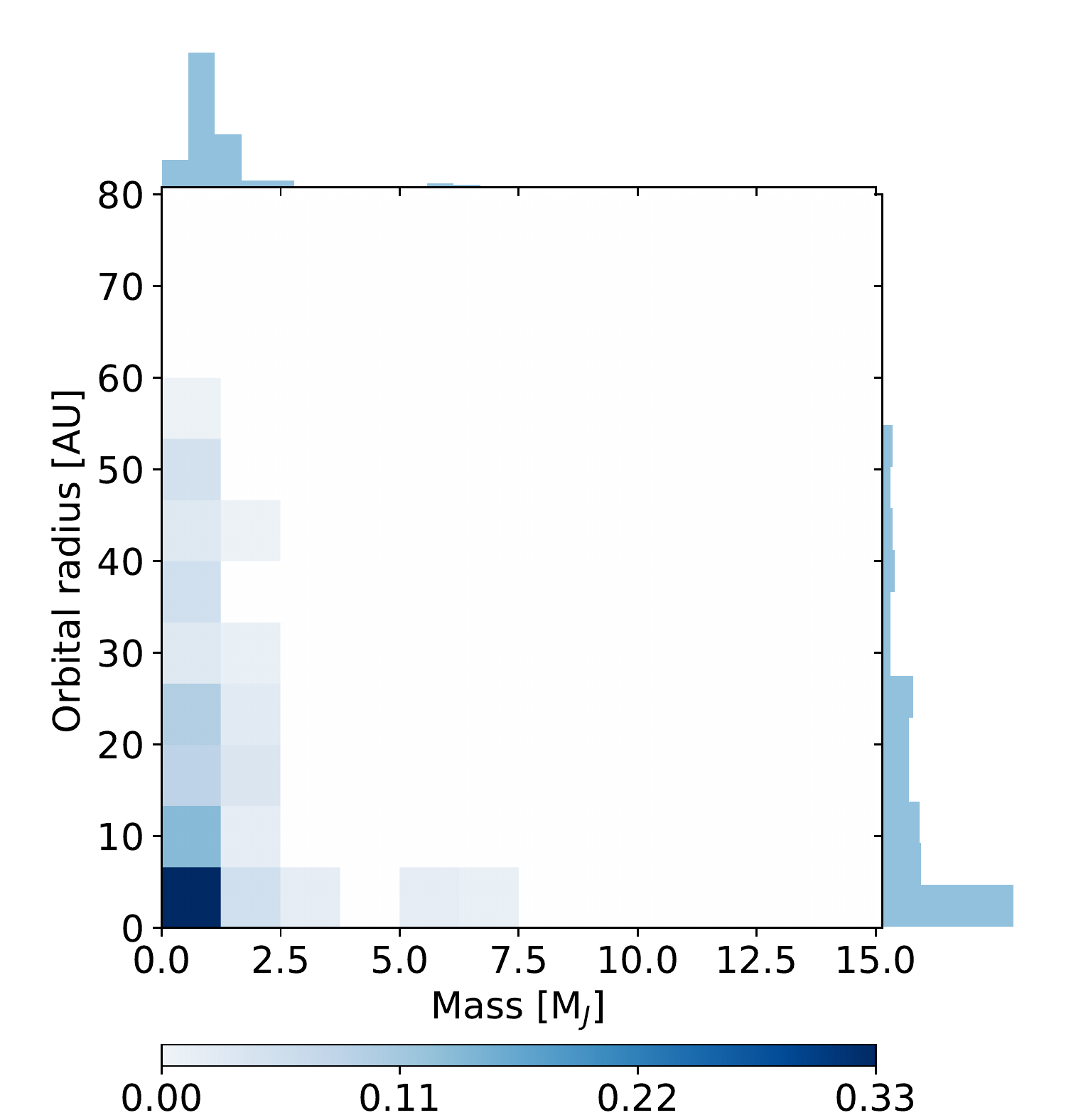} & \includegraphics[width=56mm]{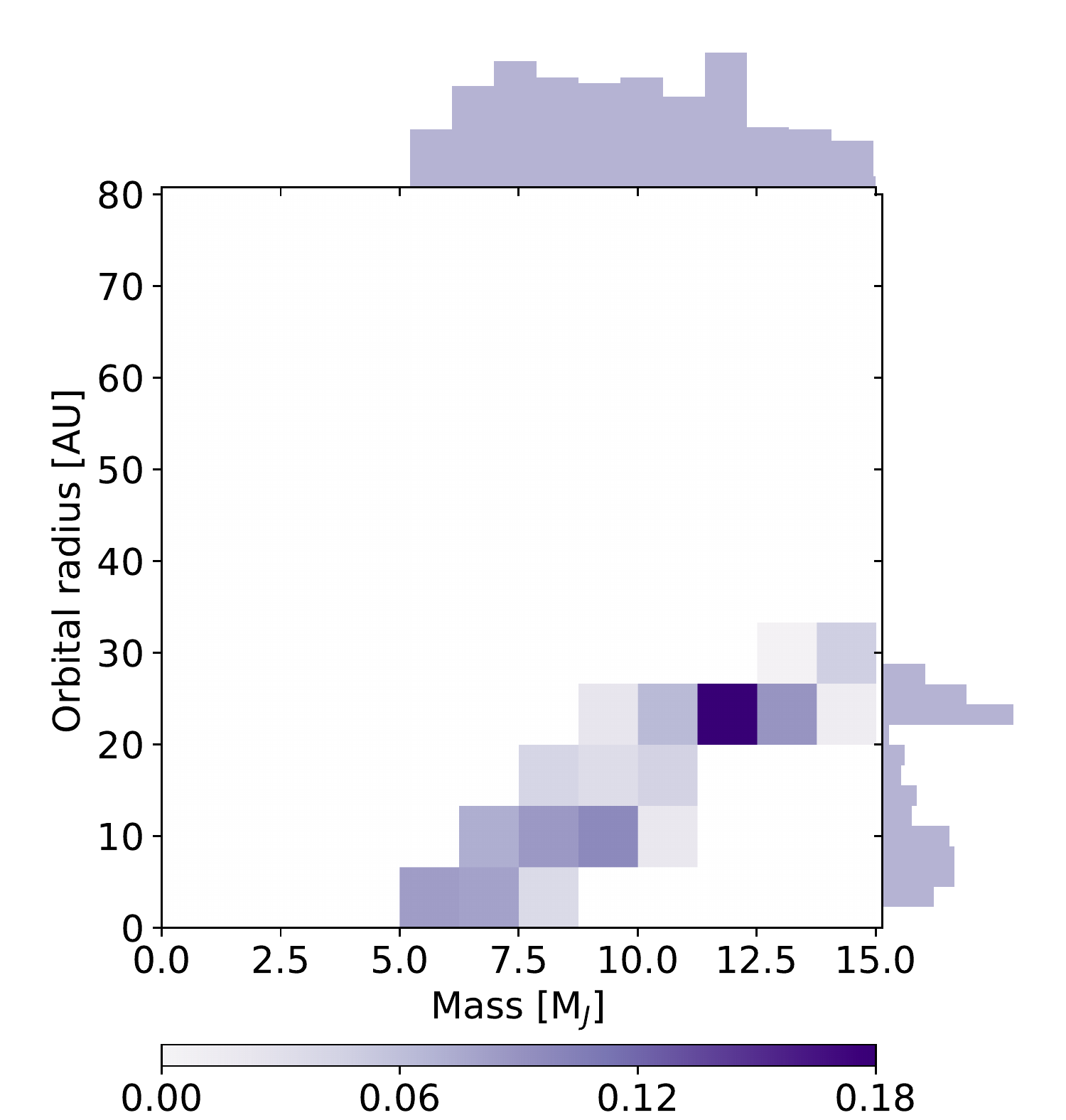} \\
	(a) IC & (b) IC-Lighter & (c) IC-Heavier\\[6pt]
	\end{tabular}
	\caption{Population of surviving clumps for different initial conditions. In IC-Lighter, the mass of all the clumps is reduced by 4 $M_{\text{J}}$ in comparison to the reference initial condition (IC). For IC-Heavier, the mass is increased by the same amount. The fragments' initial condition shapes the final population largely due to the frequent switch from type 1 to type 2 migration for massive clumps.}
	\label{fig:initial_comparison}
\end{figure}

Here we present the resulting population using the three different initial conditions discussed in section \ref{sec:initial_conditions}. The results are presented in \cref{fig:initial_comparison} and are summarized in \cref{tab:initial_conditions}. We conclude that the initial mass function of clumps has a large effect on the derived population. The heavier clumps open more gaps and undergo type 2 migration and orbit farther from the star (up to $\sim \,$30 AU), while their survival rate is high due to the very efficient gap opening. The opposite is true for a lighter initial population - lighter clumps typically follow type 1 migration and thus have a lower survival rate with most of them orbiting close to the star, within $\sim \,$15 AU.

\begin{table}[h]
	\centering
	\begin{tabular}{ccccc}
		\hline 
		\rule[-1ex]{0pt}{2.5ex} Initial condition & Gap opening & Survival & $a$ \textless 10 AU & $M$ \textgreater 10 M$_{\text{J}}$ \\ 
		\hline
		\rule[-1ex]{0pt}{2.5ex} IC & 0.08  & 0.12  & 0.52  & 0.11 \\ 
		\rule[-1ex]{0pt}{2.5ex} IC-Lighter & 0.02 &  0.17 & 0.51 & 0.00  \\  
		\rule[-1ex]{0pt}{2.5ex} IC-Heavier & 0.36  & 0.37 & 0.35 & 0.46  \\ 
		\hline 
	\end{tabular}  
	\caption{Results for different clump initial mass functions. Given are the gap opening and survival probabilities (first two columns) and frequency of clumps that orbit within 10 AU or are heavier than 10 $M_{\text{J}}$ (last two columns).}
	\label{tab:initial_conditions}
\end{table}

\subsection{Pre-collapse timescale} \label{sec:collapse_timescale}
\begin{figure}[h]
	\centering
	\begin{tabular}{ccc}
		\includegraphics[width=56mm]{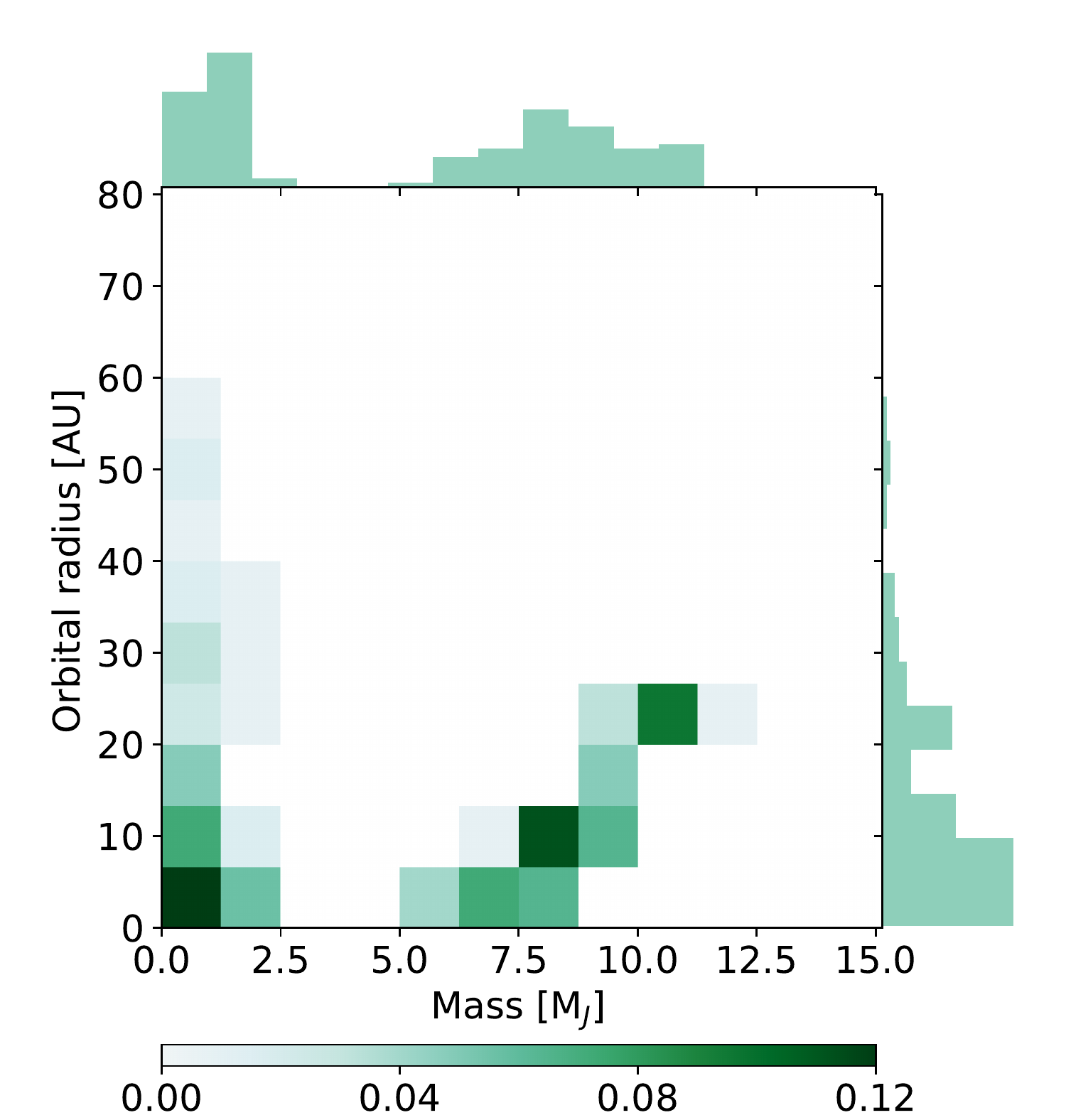} &   \includegraphics[width=56mm]{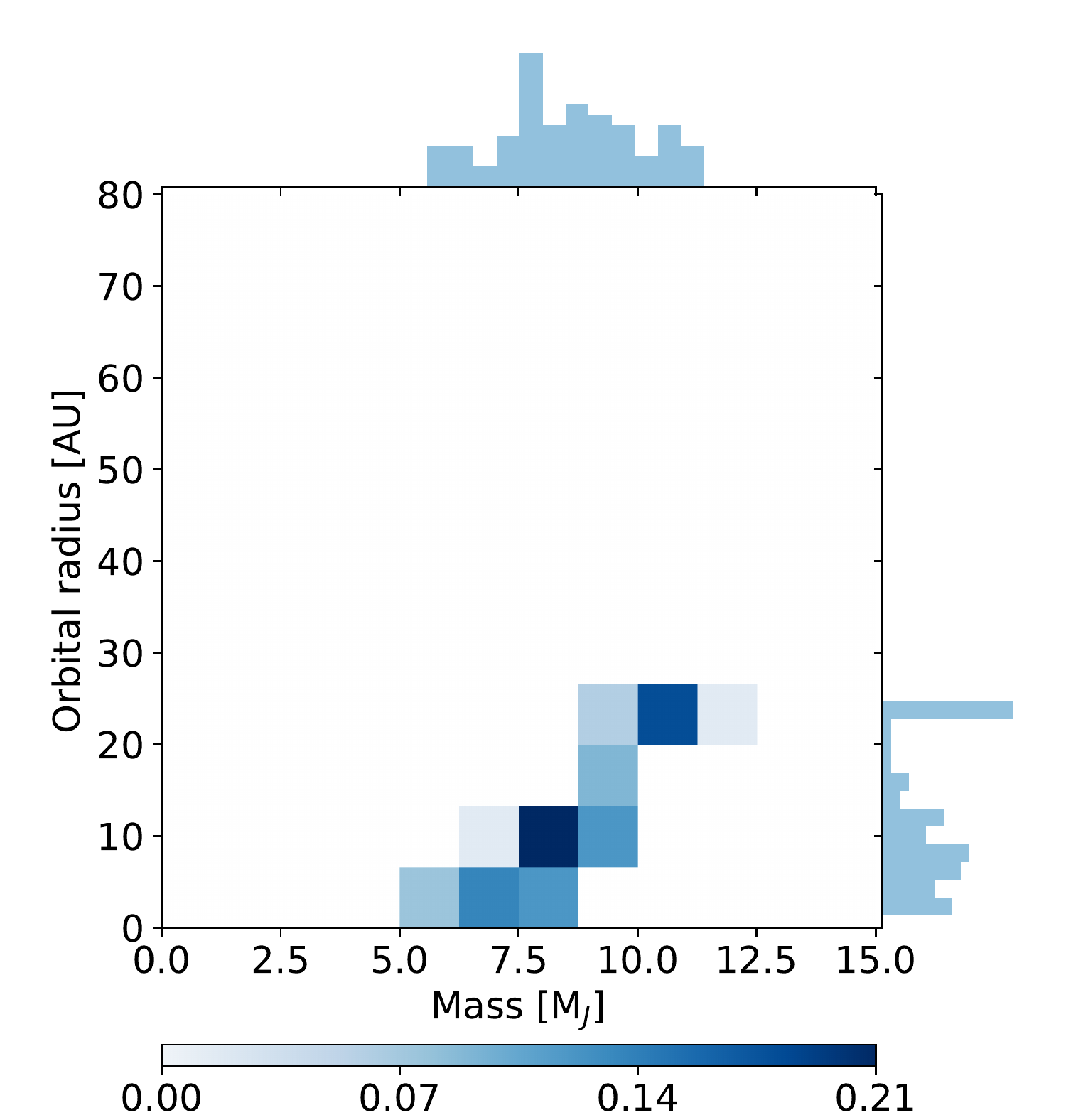} &{\includegraphics[width=56mm]{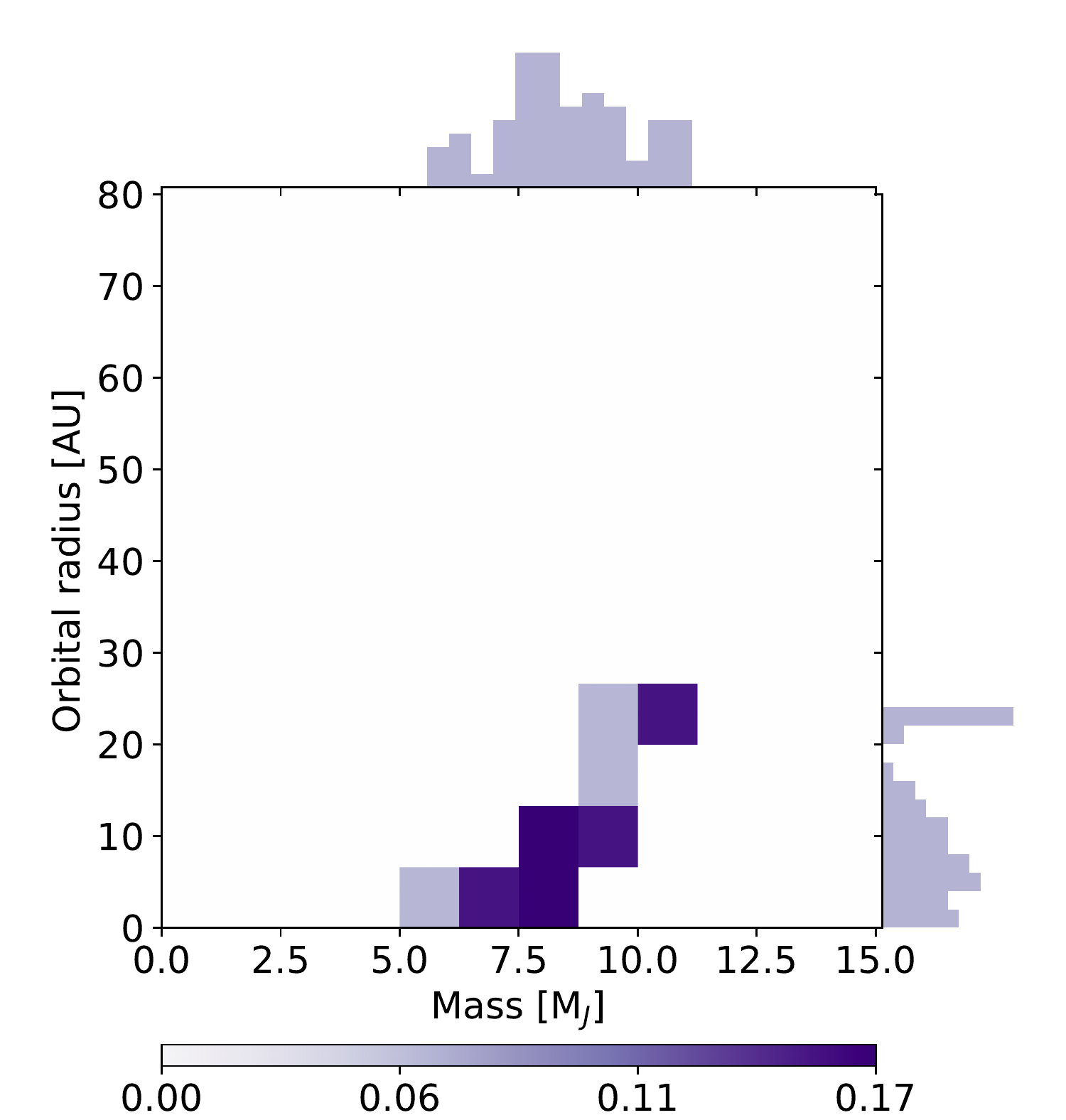}}\\
		(a) Collapse-G & (b) Collapse-H & {	(c) Constant collapse time of $10^{5}$ years}\\[6pt]
	\end{tabular}
	\caption{Population of surviving clumps for different pre-collapse timescales. For Collapse-H/C the clumps are evolved longer than for Collapse-G. In those cases, the population of lighter clumps at small orbital radii are either tidally destroyed or reach the inner disk boundary.}
	\label{fig:collapse_comparison}
\end{figure}

In this subsection we investigate the influence of the pre-collapse timescale on the results. 
Collapse-G is based on the results in \citet{Galvagni2012}, while Collapse-H uses a collapse time based on mass as described in \citet{Helled2009}. Additionally, in Collapse-C clumps evolve for $10^{5}$ years independent of their mass. We find that for the latter two cases, the survival probability is reduced to 7 \%, and that the population largely consists of massive clumps that opened a gap, which can be seen in the bottom figure in \cref{fig:collapse_comparison}. 

\begin{table}[h]
	\centering
	\begin{tabular}{ccccc}
		\hline 
		\rule[-1ex]{0pt}{2.5ex} Collapse time & Gap opening & Survival & $a$ \textless 10 AU & $M$ \textgreater 10 M$_{\text{J}}$ \\ 
		\hline
		\rule[-1ex]{0pt}{2.5ex} Collapse-G & 0.08  & 0.12  & 0.52  & 0.11 \\ 
		\rule[-1ex]{0pt}{2.5ex} Collapse-H & 0.08 &  0.07 & 0.56 & 0.16  \\  
		\rule[-1ex]{0pt}{2.5ex} Collapse-C & 0.08  & 0.07 & 0.53 & 0.19  \\ 
		\hline 
	\end{tabular}  
	\caption{The sensitivity of the inferred population to the assumed pre-collapse timescale. Gap opening and survival probabilities (first two columns) and frequency of clumps that orbit within 10 AU or are heavier than 10 $M_{\text{J}}$ (last two columns).} 
	\label{tab:collapse_time}
\end{table}

Gap opening frequencies match between all three different collapse time runs. Therefore, if the clumps manage to open a gap, they do it relatively early in their evolution within the first few thousand years. Since migration is not stopped, almost all clumps are lost due to either being tidally destroyed or reaching the inner disk boundary after $10^{5}$ years. This is evident in \cref{fig:collapse_comparison} when looking at the lower mass tail of the distribution. The difference between Collapse-H and Collapse-C is delicate: the distributions look almost identical, with the biggest difference being a slight relocation to closer orbits in Collapse-C since the clumps are allowed to evolve and migrate for longer. In all collapse-time formulations, we find no massive clumps beyond about 30 AU.

\subsection{Stellar Mass \& Disk profiles} \label{sec:disk_profiles}
While observational evidence constrains disk models, there is no universal disk model that applies to all stellar systems. In population synthesis models, a choice is made for the temperature, surface density and stellar mass that fits within these constrains but is not the only possible arrangement. In this subsection, we explore the sensitivity of the model to choosing the disk power-law surface density profile and stellar mass. 

First, we investigate the influence of stellar mass on the resulting population of protoplanetary clumps. We assume that a heavier star is also accompanied by a heavier disk and that the scaling is linear \citep{Andrews2013}. Therefore we scale the surface density profile such that the total disk mass remains at 30 \% of the stellar mass in all the cases. We create four different disk models with stellar masses of  0.6, 1.0, 1.4 and 2.4 M$_{\astrosun}$; this allows us to cover K, G, F and A-type main sequence stars.

\begin{figure}[h]
	\centering
	\begin{tabular}{cc}
		\includegraphics[width=56mm]{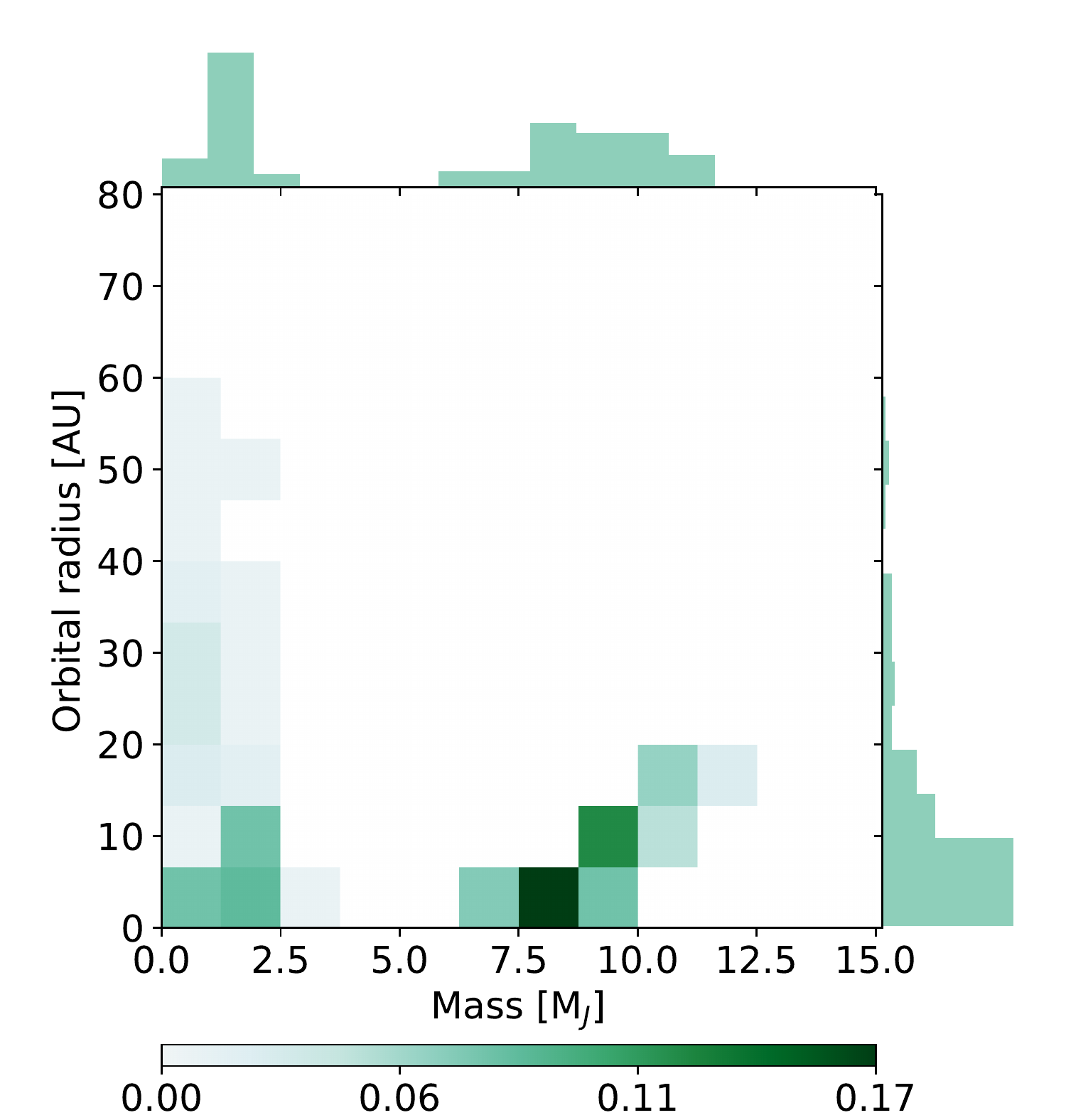} &   \includegraphics[width=56mm]{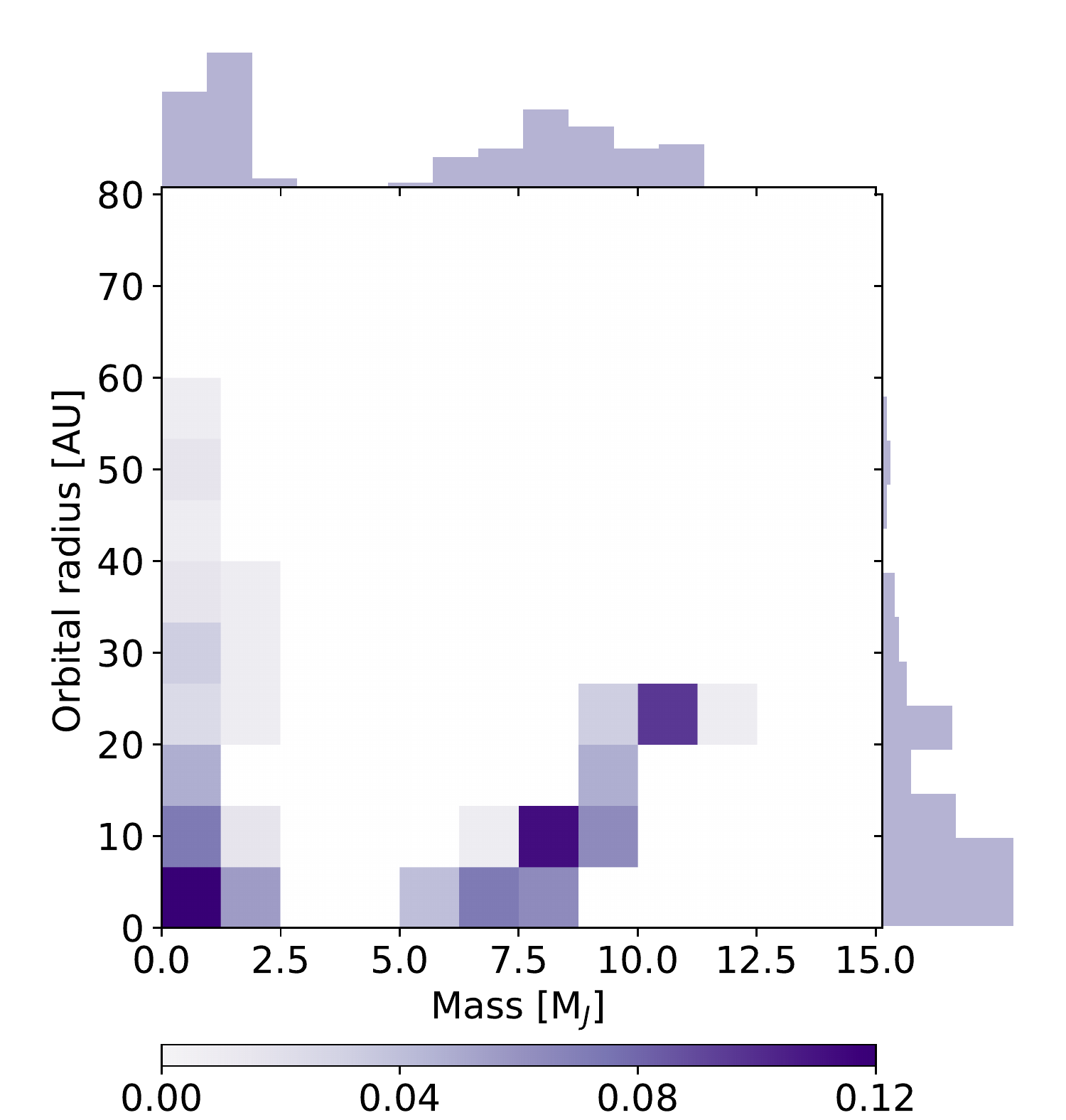} \\
		(a) 0.6 $M_{\astrosun}$ & (b) 1.0 $M_{\astrosun}$ \\[6pt]
		\includegraphics[width=56mm]{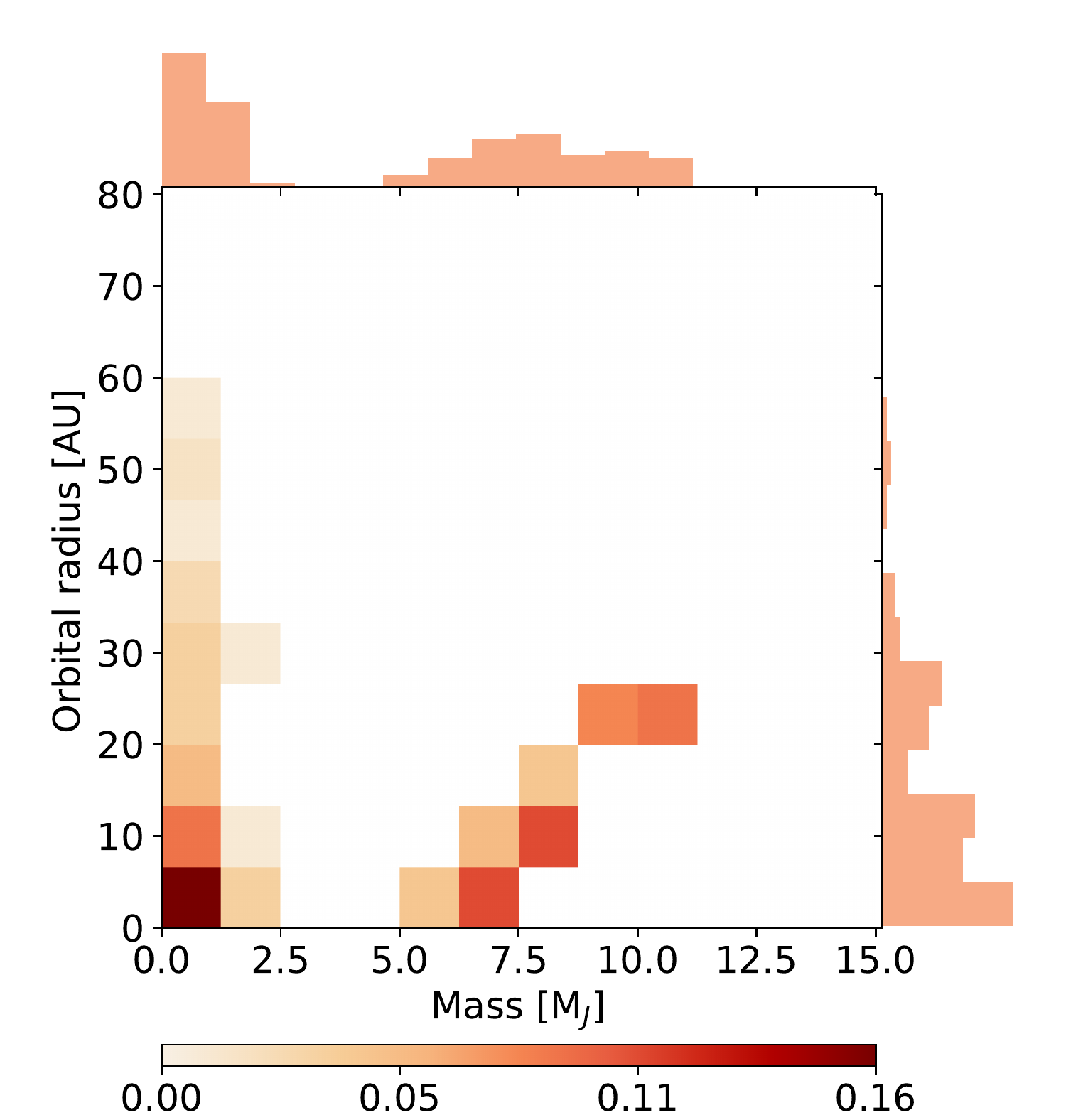} &   \includegraphics[width=56mm]{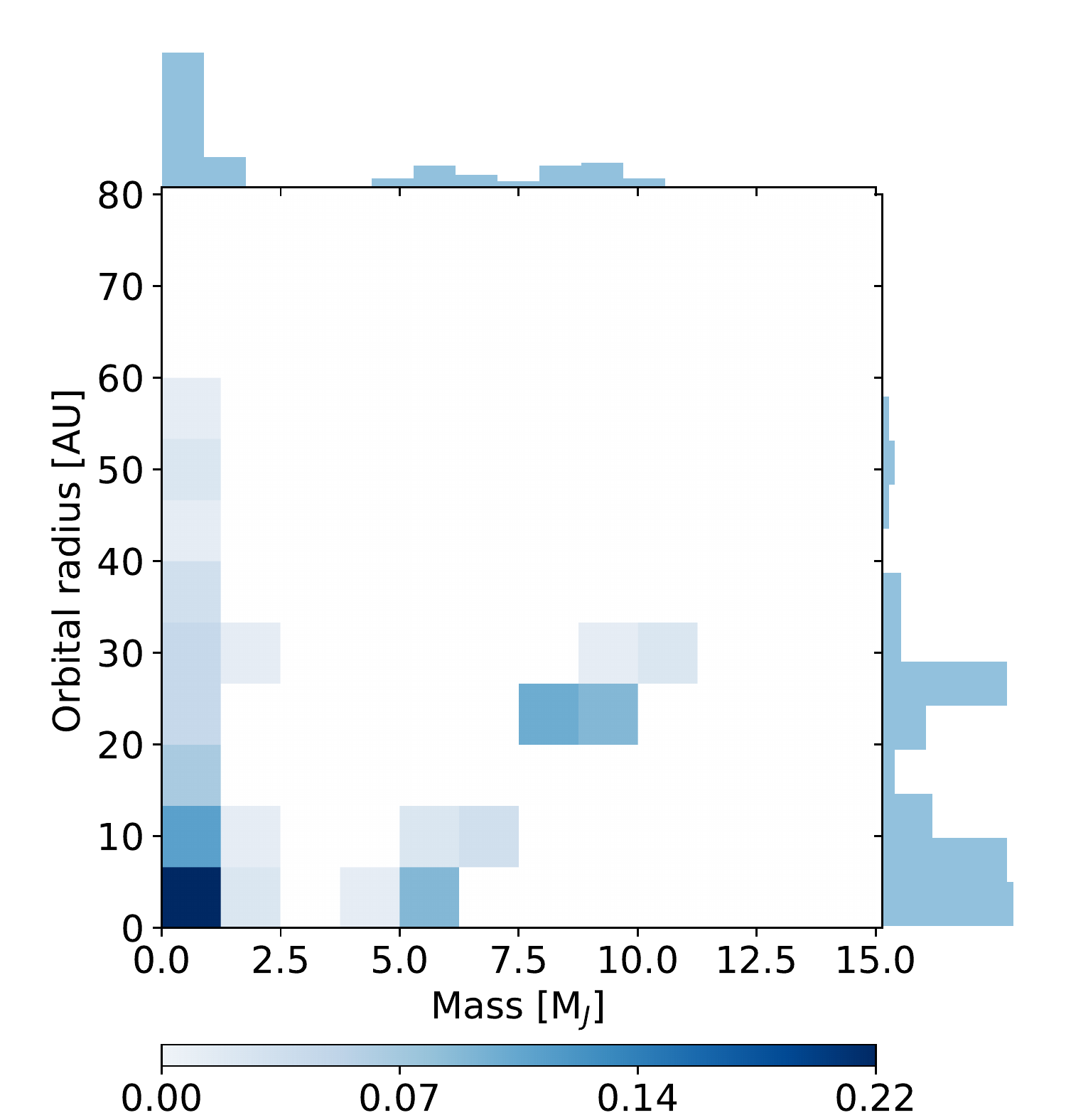} \\
		(c) 1.4 $M_{\astrosun}$ & (d) 2.4 $M_{\astrosun}$ \\[6pt]
	\end{tabular}
	\caption{Population of surviving clumps for different stellar masses ranging from 0.6 - 2.4 $M_{\astrosun}$. The faster type 1 migration caused by heavier stars causes the destruction of more and more massive clumps. Even though they are massive enough to be able to open a gap, they migrate too quickly. The population of lighter clumps orbiting close to the star is largely unaffected by stellar mass.}
	\label{fig:stellar_comparison}
\end{figure}

\begin{table}[h]
	\centering
	\begin{tabular}{ccccc}
		\hline 
		\rule[-1ex]{0pt}{2.5ex} M$_{\text{star}}$ [M$_{\astrosun}$] & Gap opening & Survival & $a$ \textless 10 AU & $M$ \textgreater 10 M$_{\text{J}}$ \\ 
		\hline
		\rule[-1ex]{0pt}{2.5ex} 0.6 & 0.08  & 0.13  & 0.68  & 0.13 \\ 
		\rule[-1ex]{0pt}{2.5ex} 1.0 & 0.08 &  0.12 & 0.52 & 0.11  \\  
		\rule[-1ex]{0pt}{2.5ex} 1.4 & 0.06  & 0.11 & 0.47 & 0.09  \\ 
		\rule[-1ex]{0pt}{2.5ex} 2.4 & 0.04 & 0.09 & 0.47  & 0.02 \\ 
		\hline 
	\end{tabular}
	\caption{Comparison of the results assuming different stellar masses. Shown are the gap opening and survival probabilities (first two columns) and frequency of clumps that orbit within 10 AU or are heavier than 10 $M_{\text{J}}$ (last two columns).}
	\label{tab:stellar_masses}
\end{table}

We find that an increased stellar mass reduces the frequency of gap opening and clump survival. Although $\tau_{\text{gap}}/\tau_{\text{cross}} \propto M_{\text{star}}^{-1/3}$ and gap opening becomes favorable, this is not enough to overcome the faster migration. We find that increasing the stellar mass shifts the population towards lighter clumps orbiting at small radial distances. In turn, a massive host star leads to less massive clumps at all radial distances.

\begin{figure}[h]
	\centering
	\begin{tabular}{ccc}
	\includegraphics[width=56mm]{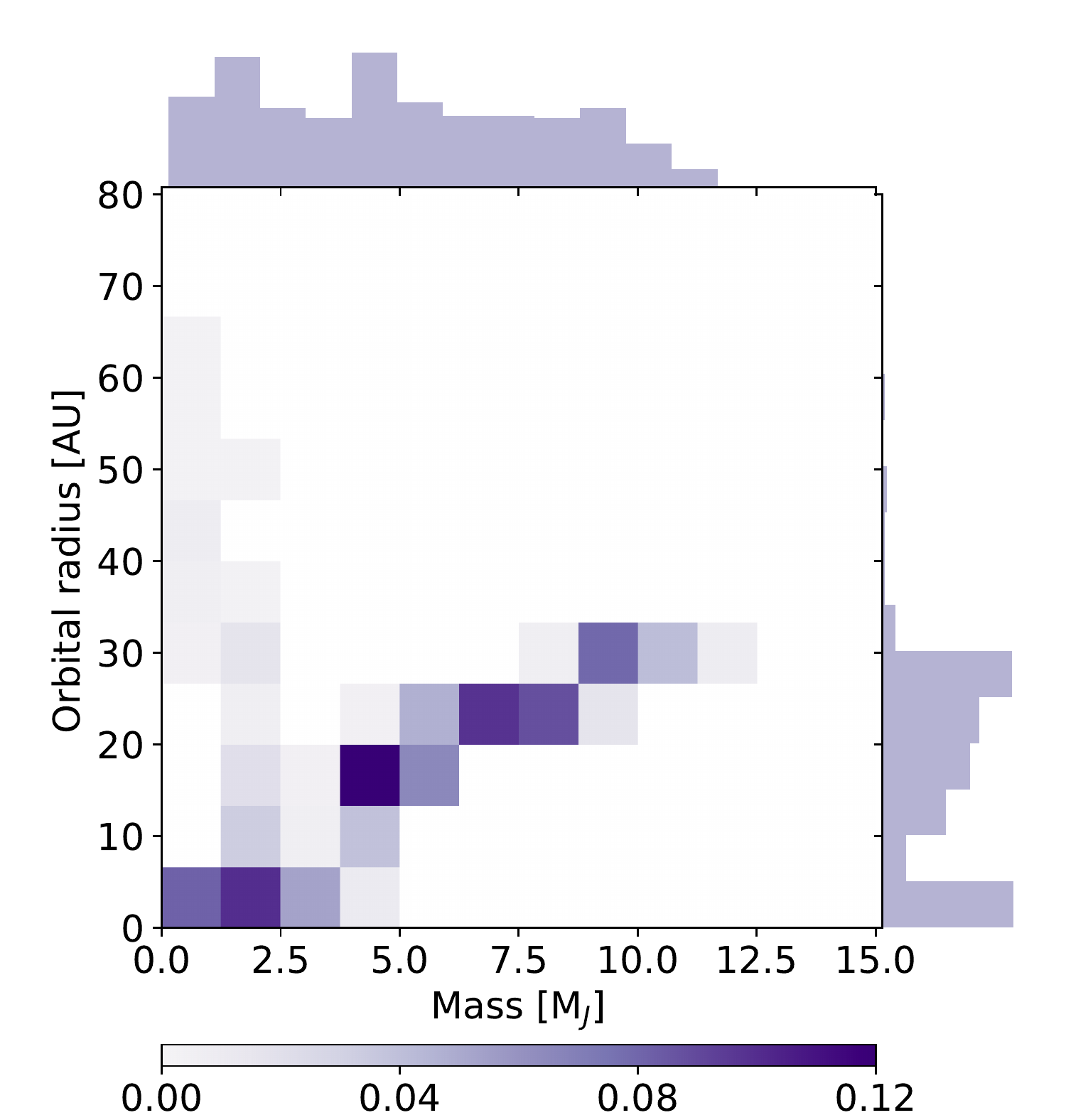} &   \includegraphics[width=56mm]{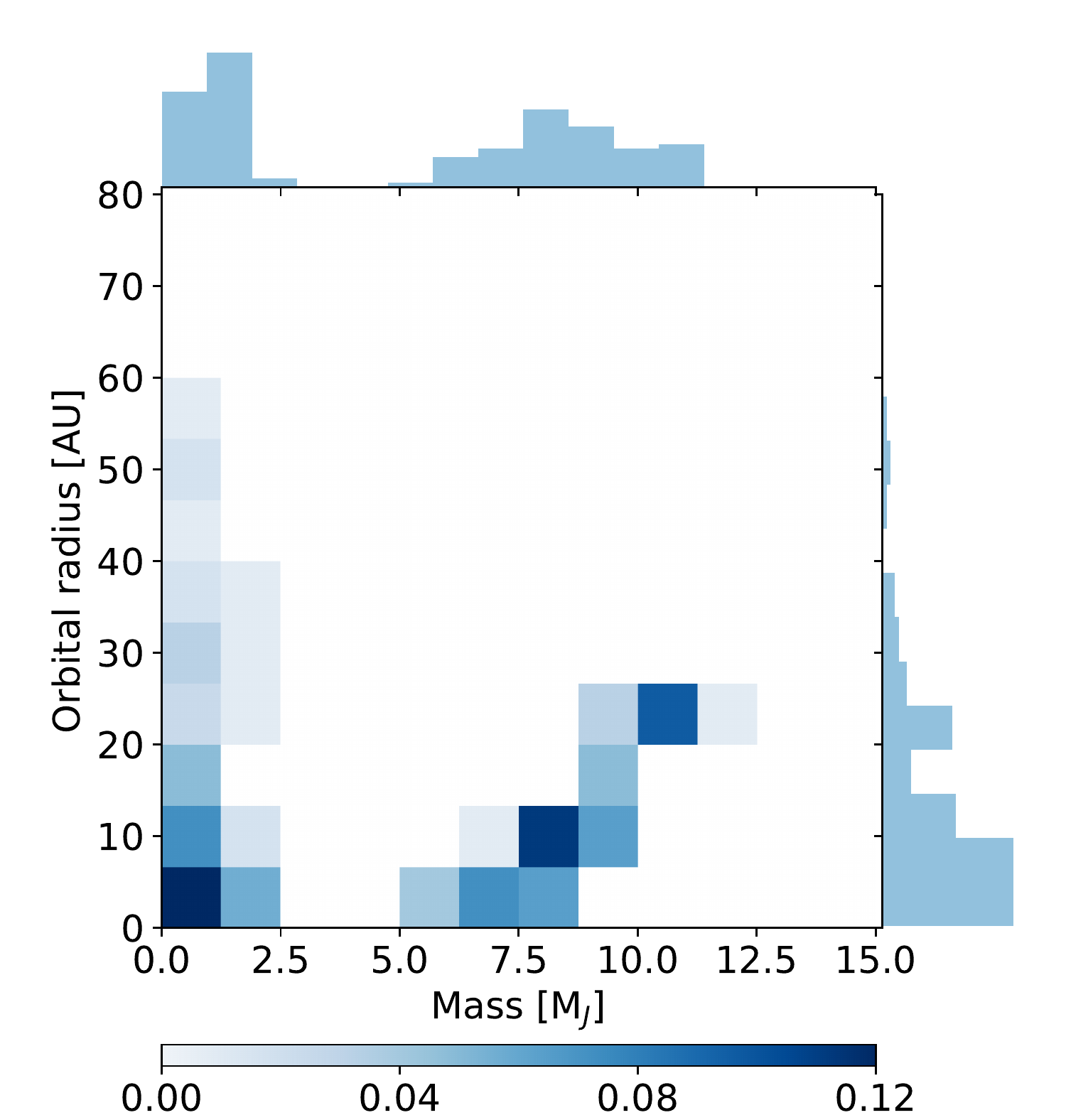} &{\includegraphics[width=56mm]{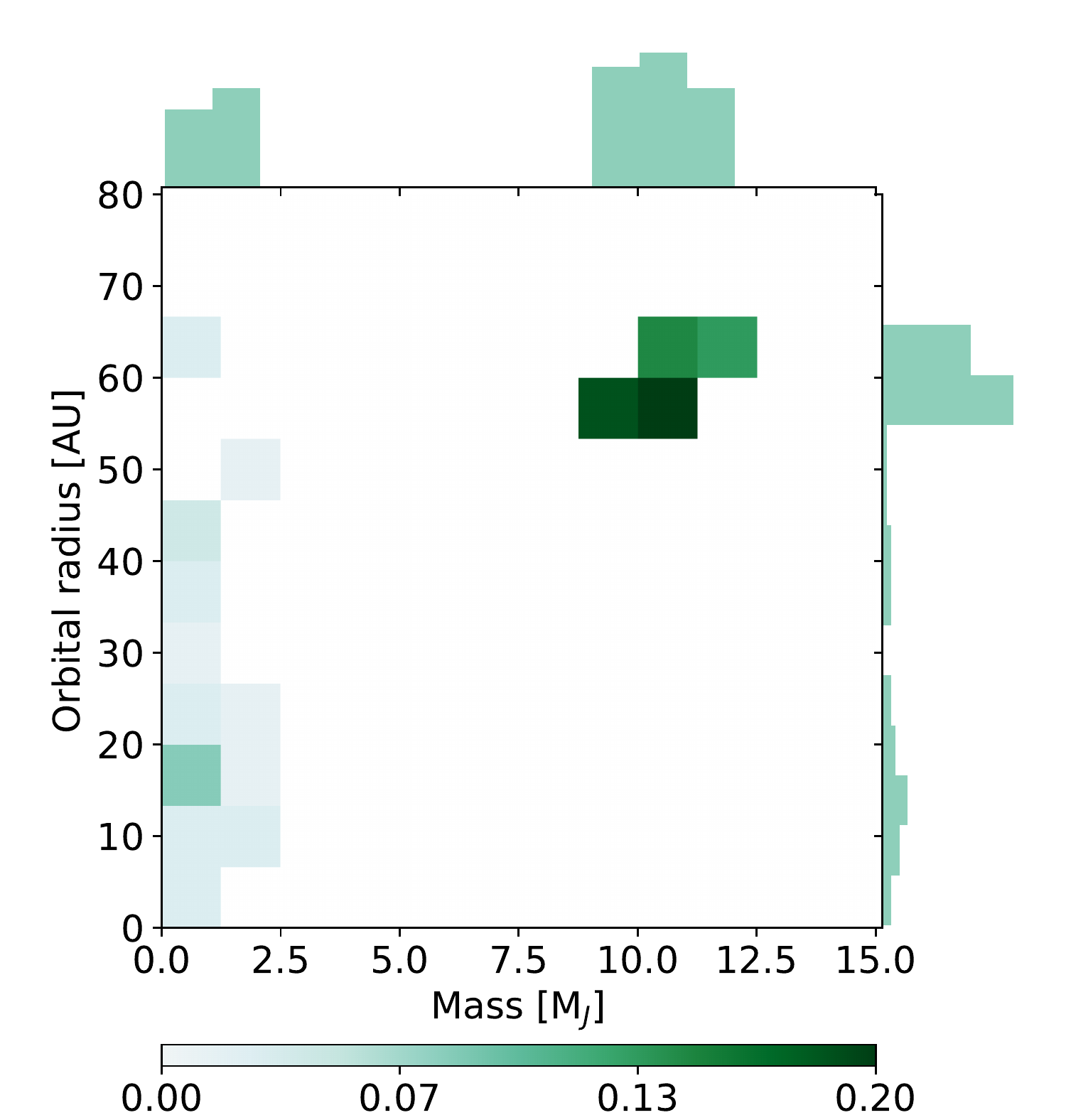} }\\
	(a) Disk-0.5 & (b) Disk-1.0 & {	(c) Disk-1.5}\\[6pt]
	\end{tabular}
	\caption{Population of surviving clumps for different disk surface density profiles. }
	\label{fig:profile_comparison}
\end{figure}

\begin{table}[h]
	\centering
	\begin{tabular}{lcccc}
		\hline 
		\rule[-1ex]{0pt}{2.5ex}  Disk profile & Gap opening & Survival & $a$ \textless 10 AU & $M$ \textgreater 10 M$_{\text{J}}$ \\ 
		\hline
		\rule[-1ex]{0pt}{2.5ex} Disk-0.5 & 0.48 &  0.28 & 0.28 & 0.05  \\  
		\rule[-1ex]{0pt}{2.5ex} Disk-1.0 & 0.08  & 0.12 & 0.52 & 0.11  \\ 
		\rule[-1ex]{0pt}{2.5ex} Disk-1.5 & 0.05 & 0.08 & 0.08  & 0.09 \\ 
		\hline 
	\end{tabular}  
	\caption{Results for the different disk surface density profiles. Again, listed are gap opening and survival probabilities (first two columns) and frequency of clumps that orbit within 10 AU or are heavier than 10 $M_{\text{J}}$ (last two columns).}
	\label{tab:profiles_comparison}
\end{table}

Most protoplanetary disk models assume a simple power-law density profile for the surface density. We now investigate the influence of the power-law on the surviving clump population. We use three disk surface densities that scale as $\Sigma \propto a^{-\sigma}$ with $\sigma = 0.5,\, 1.0,\, 1.5$. The results are presented in \cref{tab:profiles_comparison} and \cref{fig:profile_comparison}. The results show a complex relationship between the inferred population and the surface density power-law. A steeper surface density (increased $\sigma$) leads to less gap opening and lower clump survival. At the same time, more of the surviving clumps are heavier (\textgreater $10 \, M_{\text{J}}$), but also the frequency of clumps at small radial distances increases when going from $\sigma = 0.5$ to $\sigma = 1.0$. For $\sigma = 0.5$, we find a concentration of heavy clumps orbiting between 20 and 30 AU, together with a population of lighter clumps at a large variety of radial distances. The distribution in mass is rather flat. For $\sigma = 1.0$ clumps generally orbit closer to the star. The most striking difference is seen for the third profile where the survival almost mirrors the gap opening rate and many heavy clumps end up in large orbits. Using a minimum mass solar nebula (MMSN) profile causes an almost bimodal distribution compared to the previous cases, where it is more continuous. This is the only model in which we observe a population of massive clumps orbiting at large radial distances of $\sim (60 - 70)$ AU. It should be noted, however, that the MMSN density profile is inconsistent with self-gravitating disks, and typically viscous spreading flattens the surface density profile within a few dynamical timescales \citep{Durisen2006}.

\subsection{Mass accretion during gap opening}
\begin{figure}[h]
	\centering
	\begin{tabular}{cc}
		\includegraphics[width=56mm]{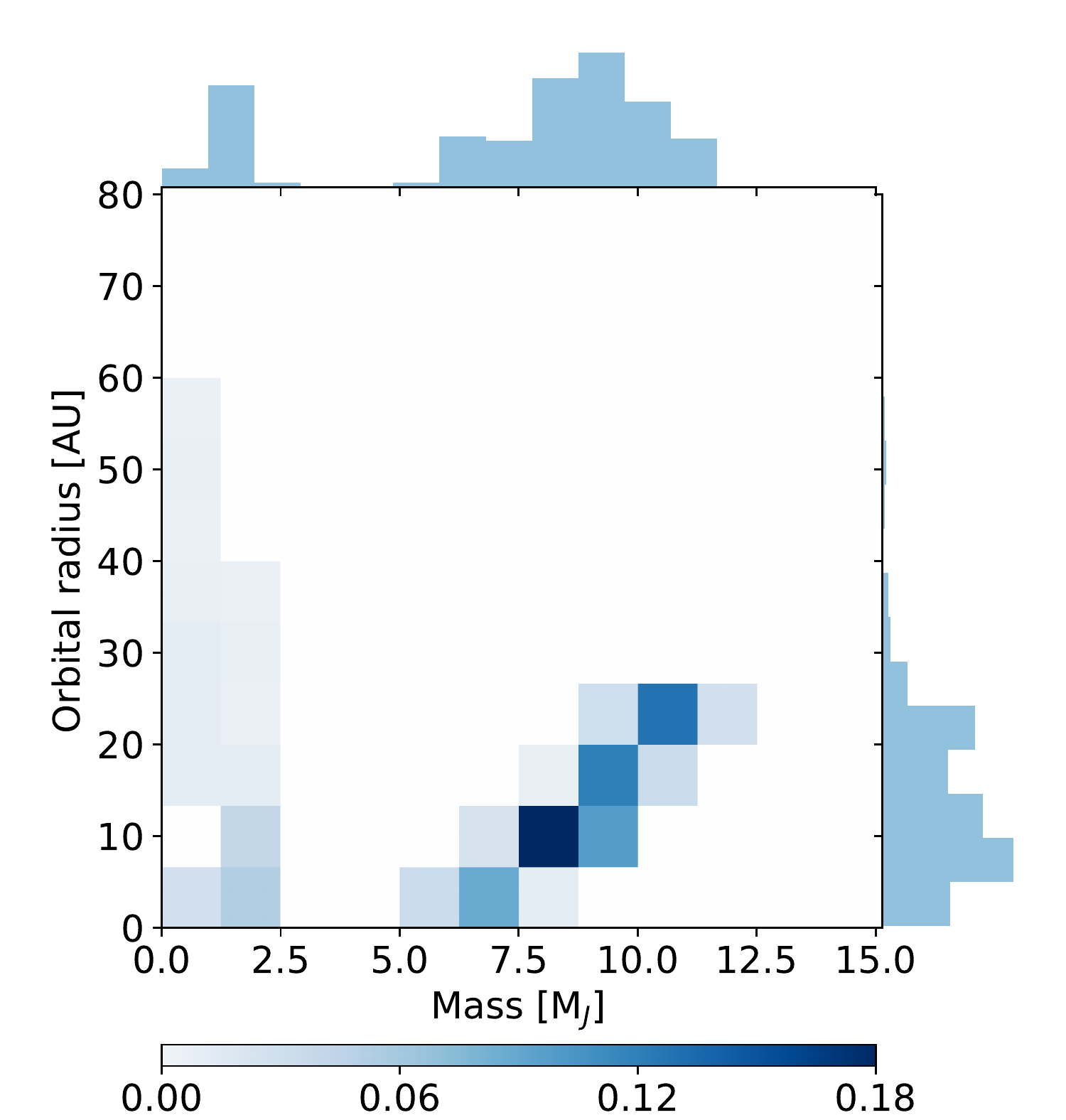} &   \includegraphics[width=56mm]{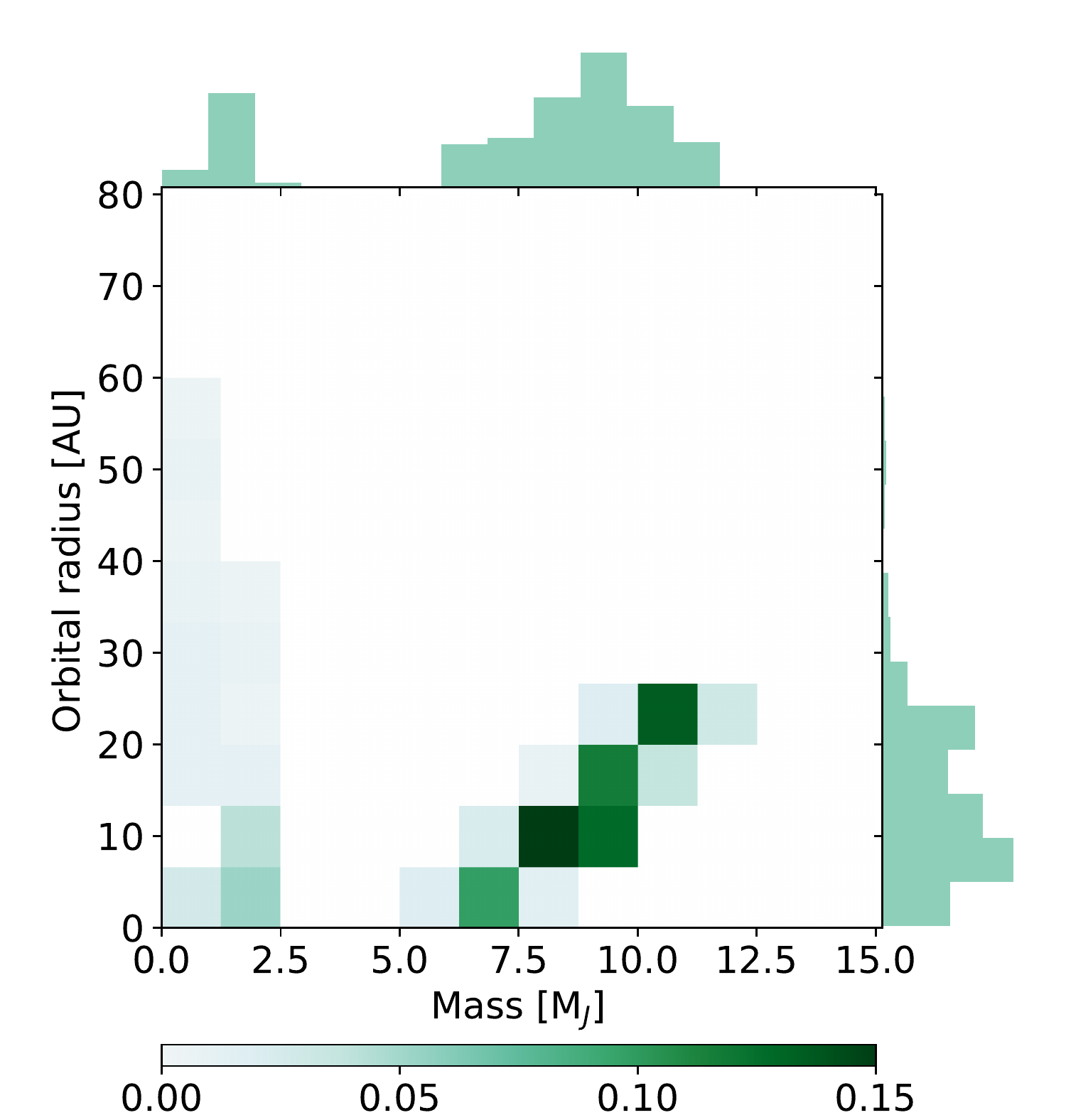} \\
		(a) $\epsilon_{\text{max}} = 0.1$ & (b) $\epsilon_{\text{max}} = 0.4$ \\[6pt]
		\includegraphics[width=56mm]{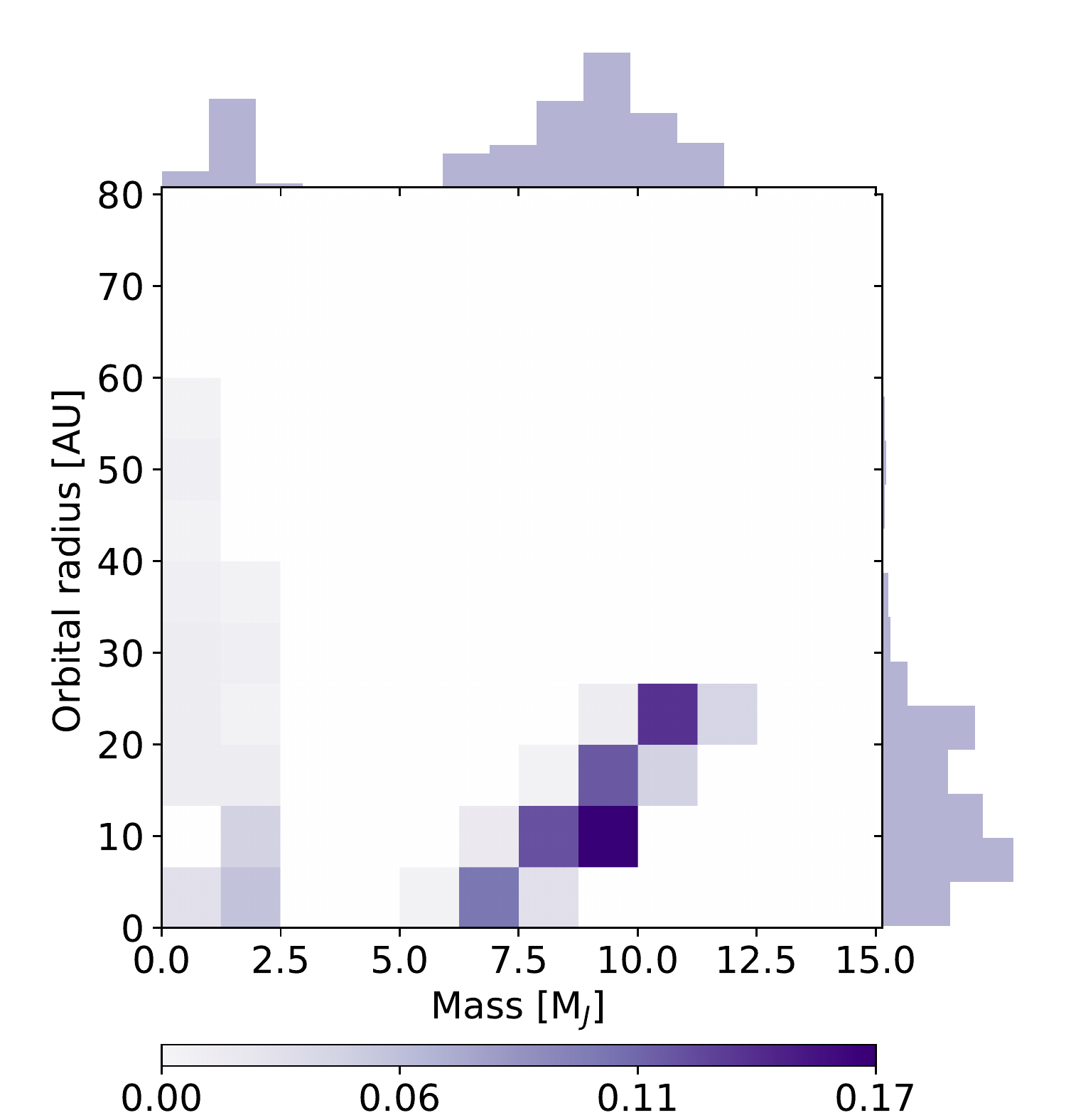} &   \includegraphics[width=56mm]{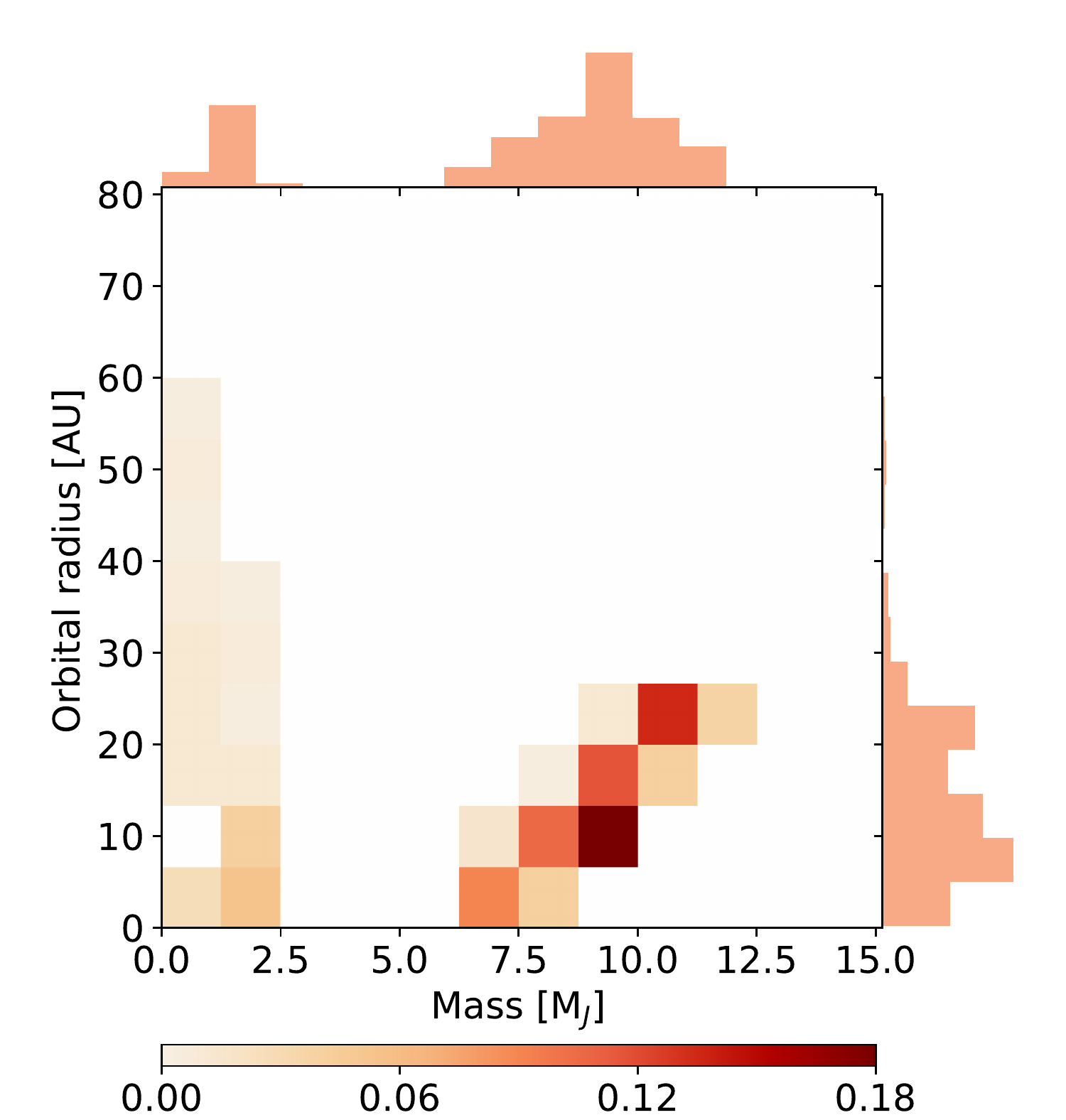} \\
		(c) $\epsilon_{\text{max}} = 0.8$ & (d) $\epsilon_{\text{max}} = 1.0$ \\[6pt]
	\end{tabular}
	\caption{Population of surviving clumps for different mass accretion efficiencies after a gap is opened. The population is not very sensitive towards a change in mass accretion. Only a minor rearrangement towards heavier clumps is observed.}
	\label{fig:epsilon_comparison}
\end{figure}

So far we have assumed that mass accretion stops once a gap is opened. However, numerical simulations suggest that tidal streams of gas that flow across the gap can lead to further mass accretion \citep{Artymowicz1996}. We therefore parametrize the accretion efficiency post gap formation using a scaling factor $\epsilon$. Simulations show a strong dependency of $\epsilon$ on the clump's mass and that it could even approach unity \citep{Lubow1999, DAngelo2002}. In order to explore the effect of the uncertainty in $\epsilon$ on the results, we use a fitting function for $\epsilon = \epsilon(M_{\text{clump}})$ \citep{Alexander2009} and set the maximum efficiency $\epsilon_{\text{max}}$=0.1, 0.4, 0.8, 1.0 for each model. The results for different maximum efficiencies are shown in \cref{fig:epsilon_comparison}. We find that the population of surviving clumps is not very sensitive to the mass accretion efficiency  after gap opening. The only difference we find in comparison to the baseline case (no accretion) is a 8-11\% increase in the frequency of surviving clumps that are heavier than $10 M_{\text{J}}$. A summary of the results is given in  \cref{tab:epsilon_comparison}. 

\clearpage

\begin{table}[h]
	\centering
	\begin{tabular}{ccccc}
		\hline 
		\rule[-1ex]{0pt}{2.5ex} $\epsilon_{\text{max}}$ & Gap opening & Survival & $a$ \textless 10 AU & $M$ \textgreater 10 M$_{\text{J}}$ \\ 
		\hline
		\rule[-1ex]{0pt}{2.5ex} 0.1 & 0.08  & 0.12  & 0.52  & 0.19 \\ 
		\rule[-1ex]{0pt}{2.5ex} 0.4 & 0.08 &  0.12 & 0.52 & 0.21  \\  
		\rule[-1ex]{0pt}{2.5ex} 0.8 & 0.08  & 0.12 & 0.52 & 0.22  \\ 
		\rule[-1ex]{0pt}{2.5ex} 1.0 & 0.08 & 0.12 & 0.52  & 0.22 \\ 
		\hline 
	\end{tabular}
	\caption{Comparison of the results using different mass accretion efficiencies $\epsilon_{\text{max}}$ after a gap is opened. Listed are the gap opening and survival probabilities (first two columns) and frequency of clumps that orbit within 10 AU or are heavier than 10 $M_{\text{J}}$ (last two columns).}
	\label{tab:epsilon_comparison}
\end{table}

\section{Comparison to previous studies and observations} \label{sec:comparison}
A population synthesis model of the disk instability scenario was developed by D.~Forgan, K.~Rice and collaborators \citep{Forgan2013,Forgan2015,Rice2015,Forgan2017}. This model includes analytical calculations of fragmentation, initial embryo mass, migration, tidal disruption, grain growth and sedimentation. The disk is evolving and photo-evaporation is included. The disk's initial surface density profile was taken to be $\Sigma \propto a^{-1}$ with an extension of 100 AU. The stellar and disk masses are both varied through the simulations and the disk viscosity is modeled using an $\alpha$ parameter that depends on the local cooling time \citep{Gammie2001}. Unlike in our work, no mass accretion after fragmentation was included and clumps were evolved for $10^{6}$ years. \citet{Forgan2015} focused on including dynamical interactions between the clumps by running N-body scattering simulations on the population derived from \citet{Forgan2013} and find an ejection rate due to fragment-fragment scattering of $\sim$25\%, which is somewhat in agreement with \citet{Terquem2002}. \\
\par 

\citet{Rice2015} investigated how perturbations from a distant stellar companion would influence the clumps. In the latest revision of their population synthesis model, \citet{Forgan2017} directly incorporate fragment-fragment interactions while the disk is still present and they include the gap opening criterion of \citet{Malik2015}. In agreement with our results, they find less efficient gap opening and rapid migration. In comparison to their previous model, they notice a small increase in lighter objects orbiting close to the star due to inward scattering and objects beyond 100 AU due to outward scattering. Nevertheless, their population is still dominated by massive giant planets and brown-dwarf sized objects orbiting at large radii. Both scattered and non-scattered models produce a population with orbital radii \textgreater 20 AU and masses \textgreater 1 $M_{\text{J}}$, which is in disagreement with our results. In addition, in many of our models, we find a significantly lower survival rate compared to the 50\% reported in \citet{Forgan2013}. 
\par

A source for the different results could be the treatment of migration; \citet{Forgan2013} use $\eta = 1$ while our baseline case is $\eta = 100$ which leads to less gap opening and a shift of the surviving fragments to lighter objects at small radial distances. Indeed, as presented in \cref{tab:eta}, using $\eta = 1-10$ leads to migration that is similar to the one derived when using only the torque balance criterion. Other possible sources for the differences could be the different clump density profiles and initial conditions. In particular, in the Forgan \& Rice model, clumps typically have masses well above 10 $M_{\text{J}}$, while our sample also includes much smaller masses. \\
\par

Another recent and actively evolving population synthesis model has been presented by S.~Nayakshin and collaborators \citep{Nayakshin2015b, Nayakshin2015c, Nayakshin2016}. Unlike the \citet{Forgan2013} model, it includes solids (pebbles) accretion onto protoplanetary clumps. The assumed disk model is similar to that of \citet{Forgan2013} with a star of 1 M$_{\odot}$, and include planet-disk interaction as described in \citet{Nayakshin2012}. This allows a self-consistent description of gap opening and gas deposition from tidally disrupted embryos at a given location in the disk. The gap opening criterion is that of \citet{Crida2006} and the internal structure and evolution of the protoplanets are modeled numerically rather than analytically. Our power-law density profiles are consistent with those of \citet{Nayakshin2015b}. Like in our model, clumps evolve in isolation, which is a simplifying assumption (see e.g. \citet{Vorobyov2006, Boley2011}). \\
\par

The results of the \citet{Nayakshin2015b} model can be summarized as follows: the inferred population mostly consists of small-core-dominated planets. There is a sharp drop in frequency of planets more massive than 0.1 $M_{\text{J}}$, which we do not observe in our results. Generally, in agreement with our findings, the simulated protoplanets cover a wide variety in orbital radii, although we find a smaller number of planets on wide orbits. Again, a direct comparison is difficult due to the many differences, not limited to but including longer evolution time, different disk models, $\alpha$ parameter, gap opening and thus migration. The reason that heavier clumps survive could be a result of more efficient gap openings and the inclusion of solids in the simulated internal structure (e.g., a core). This could make fragments  more protected from stellar tidal disruption. \\
\par 

The large number of detected exoplanets makes it possible to compare the synthetic populations with the observed one. 
It is important to note, that imaging surveys target massive
planets around young stars, but these young stars can still be several tens of Myr old. Therefore a direct comparison with the models is challenging as significant dynamical and planetary evolution could affect the final population. This is however an issue with all current population synthesis models. \\
\par

\citet{Vigan2017} combined the results of several imaging surveys and then compared the results with the disk instability population synthesis model of \citet{Forgan2013}. This model produces populations with properties that overlap with core accretion but they also produce a large population of massive objects at large radial distances. \citet{Vigan2017} estimate that a significant fraction of planets formed in the population instability model by Forgan \& Rice could potentially be detected. However, since these detectable high mass planets orbiting at large radial distances have not been detected, Vigan et al. concluded that disk instability is rare. \\

A comparison of the \citet{Nayakshin2016} model with observations is presented in \citet{Nayakshin2017a}. Both the synthetic and the observed populations are dominated by small planets. The abundance of planets more massive than $\sim 0.1 M_{\text{J}}$ is low and gas giants between 0.1 and 1 AU are rare. The simulated planets cover a large range of radial distances. In the observed population, lower-mass planets (below $\sim 1 M_{\text{J}}$) at intermediate orbital radii are absent, which is most likely due to observational biases. There is a discrepancy between the frequency of massive planets at $\sim$10 AU between the model and observations. The model also fails to reproduce the population of small planets below 0.1 AU. \\

A recent study of the effect of disk cooling on the formation of planets in the disk instability model is presented in \citet{Boss2017}. While this is not a rigorous population synthesis study, the author investigates several disk models using the $\beta$ cooling approximation. It was found that a low initial minimum Toomre $Q_{\text{i}}$ allows for disk fragmentation and clump formation at small radial distances, from 4 to 20 AU. \citet{Boss2017} presents a comparison between observations and the model results, showing an agreement for radial distances below 6 AU.The models also predict a significant number of gas giants with masses of $\sim 1 M_{\text{J}}$ at radial distances between 6 and 16 AU. \\

Here we find that different model assumptions result in very different populations. In particular, our results show that in the disk instability model, giant planets at large radial distances are rare. We therefore urge for caution when drawing conclusions that come from comparing population synthesis models, and planet formation models in general, with observations. In addition, it is very clear that model assumptions strongly affect the inferred population and that a large diversity in orbital radius and mass of giant planets in the disk instability model can be produced.

\section{Discussion and Conclusions} \label{sec:discussion}
\begin{figure}[h]
	\centering
	\includegraphics[scale=0.45]{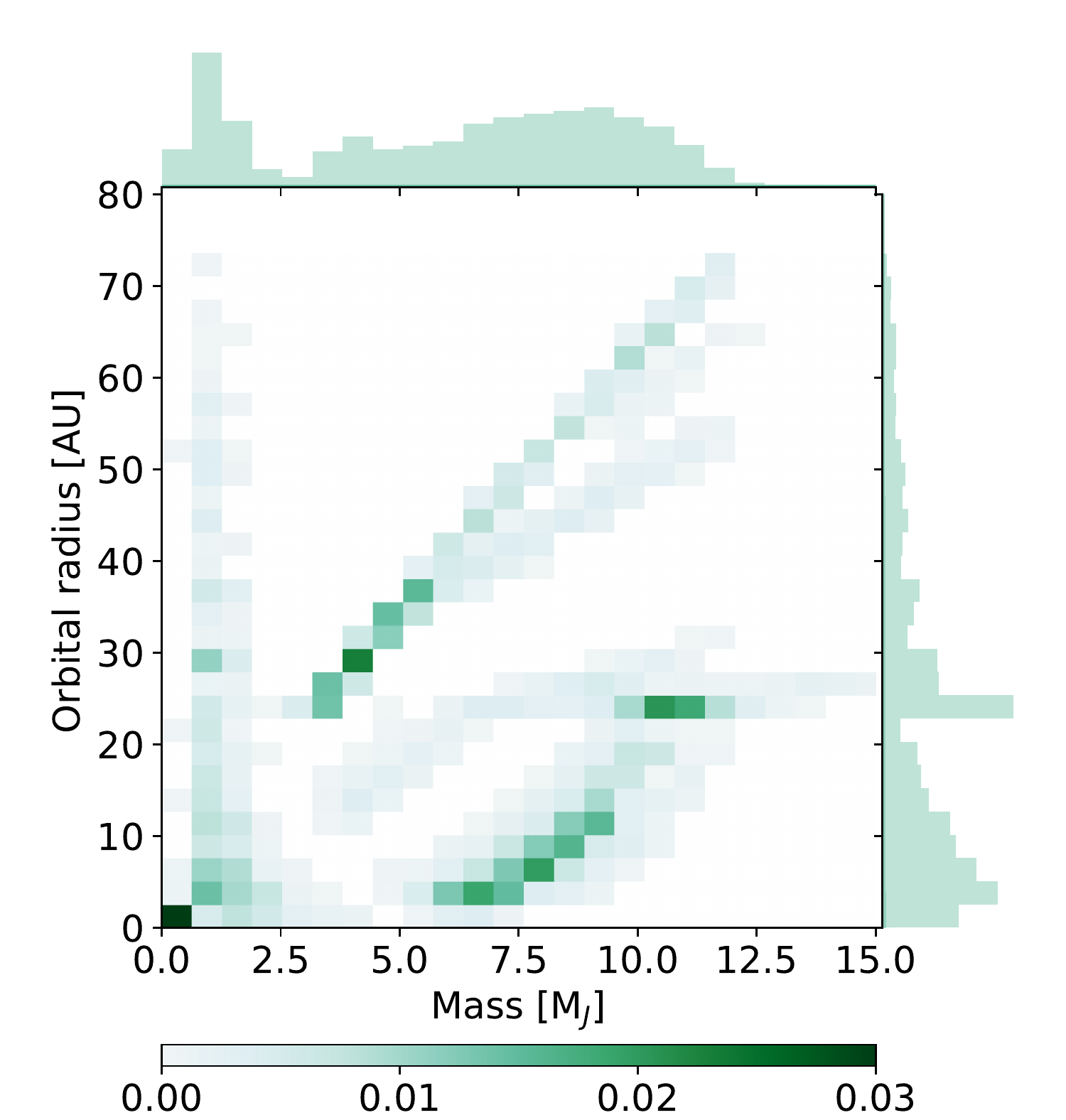}
	\caption{Derived population of surviving clumps combined from all the models. The distribution in mass is mostly flat with a peak at $\sim 1 M_{\text{J}}$. For the radial distance, we find that clumps typically have orbital radii \textless 30 AU. Note that in this figure all the models have equal weights while in reality some of them are less realistic than others. 
	Therefore this combined population should not be directly compared to observations without further analysis. Nevertheless, this figure clearly demonstrates the diversity of clumps in the disk instability scenario. }
	\label{fig:hist_full}
\end{figure}

We present a new population synthesis model of giant planets in the disk instability scenario. In particular, we explore the influence of the model assumptions and parameters such as clump density profiles, initial conditions, collapse time, disk profiles, migration and mass accretion on the inferred population. Figure \ref{fig:hist_full} combines the results of all the models we consider in one histogram. The clumps occupy a large range of masses and radial distances although most of the population is orbiting below 40 AU and there is a peak in the mass distribution at around $1 \, M_{\text{J}}$. It should be noted, however, that our simulations stop when clumps reach dynamical collapse and therefore this population might not represent the final population of planetary clumps since mass accretion and migration may still have a significant influence. \\

We investigate the effect of the gap opening criteria on migration. We find that when using $\alpha = 0.05$ no gaps are opened since the torque balance criterion is not satisfied. 
We therefore use the standard $\alpha = 0.005$ which allows for gap opening. 
In principle, this low $\alpha$ value is well below the typical values of 0.01-0.05 found in simulations of moderately massive, marginally stable self-gravitating disks for which a local $\alpha$-disk prescription is meaningful \citep{Lodato2005}. However, our choice of $\alpha$ is conservative for two reasons. First, with time, as gas is accreted onto the central star (and clumps) the disk surface density drops and therefore the disk becomes more stable and less gravito-turbulent. If the magneto-rotational instability (MRI) develops and renders it turbulent again, lower values of $\alpha$ are expected \citep{Fromang2007}. Second, with a larger $\alpha$ value migration is even faster and no gaps are opened, strengthening our point that massive gas giants at large radial distances ($>$ 50 AU) are rare. \\
\par 

Furthermore we find that gap opening is rare when using the criterion based on the comparison between the gap opening and crossing timescales. This significantly alters the final distribution of clumps and their survival probability: clumps with masses above $10 \, M_{\text{J}}$ rarely survive and the resulting population is characterized by lower masses and closer orbits. Our baseline simulations use a scaling factor for the gap opening timescale of $\eta = 100$, and we find that for $\eta = 10$ the results are similar to the ones derived when only considering torque balance. Our produced populations are not very sensitive towards a change in the mass accretion efficiency during gap opening. The distributions are almost the same, with only a minor re-arrangement towards higher masses. We find at most an increase of 11\% of surviving clumps that are heavier than $10 \, M_{\text{J}}$. \\

We introduce a realistic power-law density profile for clumps that matches high-resolution studies \citep{Galvagni2012,Szulagyi2016} and detailed clump evolution models \citep{Helled2010,Vazan2012}. Such a density profile increases both survival and gap opening probabilities compared to the quasi-homogeneous case. This alters the population of surviving clumps in our reference model Disk-1.0-Gap: the power-law density profile produces a clump population with a large mass range and orbital distances mostly below 30 AU. In contrast, quasi-homogeneous clumps lead to a population that is dominated by low mass clumps. We find that clumps with realistic density profile must become denser by several orders of magnitude as they evolve in order to survive tidal downsizing. \\

We explore the effect of the initial clump mass on the derived clump population and find that the fragments' initial condition shapes the final population due to the more common type 2 migration for massive clumps. The population from lower-mass primordial clumps is mostly dominated by sub-3$\, M_{\text{J}}$ at varying radial distances. For the heavier initial condition, we find surviving clumps of up to $15 \, M_{\text{J}}$ at radial distances below 40 AU. We therefore conclude that robust predictions of clump masses can be made only when we have a better understanding of the initial mass function of planetary clumps. In addition, we find that the pre-collapse timescale also plays a crucial role in determining the final population. Evolving clumps for up to $10^{5}$ years results in no low-mass clumps since they tend to either be tidally destroyed or reach the inner disk boundary. In those cases, the population consists of massive clumps (beyond $5 \, M_{\text{J}}$) at radial distances up to 30 AU. \\

We also check the sensitivity of the results to the assumed disk surface density profile and stellar mass. An increase in the stellar mass causes faster type 1 migration and consequently the destruction of massive clumps. Although these clumps would be massive enough to open a gap by torque balance, they migrate too quickly. The lighter clumps are primarily unaffected by the stellar mass. Regarding the disk surface density, we find that a steep profile decreases the gap opening frequency and leads to lower clump survival rates. Increasing $\sigma$ from 0.5 to 1.0 mostly affects the mass distribution of the surviving clumps, going from a nearly uniform distribution to a bimodal one. Disks with $\sigma = 1.5$ result in a very different population, and represents the only case in which the inferred population is dominated by massive clumps at large radial distances. \\

Clearly more work is needed. First, in our models the clump evolution is stopped at dynamical collapse and it is desirable to model the long-term evolution using planetary evolution models. Second, to describe the disk viscosity we use the simple $\alpha$ parameter description while it is unclear whether a constant $\alpha$ is appropriate for gravitationally unstable disks, and how it changes with the disk's  properties. In addition, the way type 2 migration is implemented in this model is binary: either a planet opens a gap and evolves on the viscous time scale or it does not. In reality, as suggested by numerical simulations \citep{Crida2006,Crida2007,Fung2016}, there are varying depths of gaps, which strongly affects the planetary migration. Our models use the simplifying assumption of a static disk with only inward migration. More detailed 3D hydrodynamical simulations of evolving disks suggest that outward migration is also expected \citep{Boss2013}. Therefore it is desirable to include disk evolution and its effect on the resulting population self-consistently. Furthermore, our model does not include grain growth or planetesimal/pebble accretion and thus cannot address core-dominated planets. Finally, clump interactions and disk feedback should also be considered. N-body simulations show that clump-clump scattering can affect the population \citep{Forgan2015,Terquem2002} and lead to clump ejection. \\

In this study, we show that the resulting clump population strongly depends on the assumed disk model, internal structure, collapse time and initial conditions. Varying these parameters drastically influenced gap opening and survival probability, and the physical properties of the surviving clumps.  Although our results display this substantial diversity, there are some robust trends:

\begin{itemize}
	\item An additional gap opening criterion causes rapid migration and less efficient gap opening.
	\item Changing the model parameters such as stellar mass, disk density profile, initial mass function of clumps and their density profiles leads to significant changes in the surviving clump population.
	\item Protoplanets occupy a mass range between 0.01 and 16 $M_{\text{J}}$ and can orbit very close to the central star or as far out as 75 AU.
	\item Most massive 	protoplanets have orbital radii of 10--30 AU.
	\item All the investigated cases suggest that massive giant planets at very large radial distances are rare in agreement with observations. 
\end{itemize}

This study represents a step towards a better understanding of the resulting population of clumps in the disk instability model. We suggest that future population synthesis models should consider a large parameter space and that comparison with observations should be done with great caution. Finally, we argue that there is much room for improvements in planet population studies of the disk instability model, and we hope that future work will lead to a more coherent understanding of the predictions of this formation scenario. 

\newpage
\bibliography{population_synthesis_simon}
\bibliographystyle{aasjournal}
  
\end{document}